%% file: JBSMSE-961.tex
\documentclass[twocolumn]{svjour3} 

\usepackage[american]{babel}
\usepackage[utf8]{inputenc}
\usepackage[T1]{fontenc}
\usepackage{ae}
\usepackage{aecompl}

\usepackage{commath}

\usepackage[numbers]{natbib}
\usepackage{graphicx}
\usepackage{subfigure}
\usepackage{booktabs}
\usepackage{color}
\usepackage{bm}

\input{macros}

\journalname{J Braz. Soc. Mech. Sci. Eng.}

\begin{document}

\title{Assessment of a transient homogeneous reactor through in situ adaptive tabulation}

\author{Americo Cunha~Jr \and \mbox{Lu\'{i}s Fernando Figueira da Silva}}

\institute{A. Cunha~Jr (corresponding author) \and L. F. Figueira da Silva \at
              Department of Mechanical Engineering\\
              Pontif\'{\i}cia Universidade Cat\'{o}lica do Rio de Janeiro\\
			  Rua Marqu\^{e}s de S\~{a}o Vicente, 225 - G\'{a}vea\\
			  Rio de Janeiro, 22451-900, Brasil\\
              \email{americo.cunhajr@gmail.com}
}

\date{Received: date / Accepted: date}

\maketitle

\begin{abstract}
The development of computational models for the 
numerical simulation of chemically reacting flows operating in the turbulent 
regime requires the solution of partial differential equations that represent the balance 
of mass, linear momentum, chemical species, and energy. 
The chemical reactions of the model may involve detailed reaction
mechanisms for the description of the physicochemical phenomena. 
One of the biggest challenges is the stiffness of the numerical simulation 
of these models and the nonlinear nature of species rate of reaction.
This work presents a study of \emph{in~situ} adaptive tabulation (ISAT)
technique, focusing on the accuracy, efficiency, and memory 
usage in the simulation of homogeneous stirred reactor models using
simple and complex reaction mechanisms. The combustion of 
carbon monoxide with oxygen and methane with air
mixtures are considered, using detailed reaction mechanisms 
with 4 and 53 species, 3 and 325 reactions respectively.
The results of these simulations indicate that the developed 
implementation of ISAT technique has a absolute global error 
smaller than 1\%. Moreover, ISAT technique provides gains, 
in terms of computational time, of up to 80\% when compared 
to the direct integration of the full chemical kinetics. However, 
in terms of memory usage the present implementation of ISAT 
technique is found to be excessively demanding.

\keywords{thermochemistry reduction \and in situ adaptive tabulation \and 
stirred reactor simulation \and detailed reaction mechanism}
\end{abstract}

\section{Introduction}

Combustion is a physicochemical phenomena characterized by a
sequence of chemical reactions (most of then exothermic) which
converts fuel/oxidizer mixtures (fresh gases) into combustion 
products (burned gases). Due to the large amount of energy released
in these exothermic reactions, the combustion process is widely
used in the operation of industrial devices such as gas turbines,
process furnaces, etc.

Design and optimization of devices which operate based on
combustion processes is an important engineering task,
which demands the development of computational models
as predictive tools. Such computational models may
require the solution of partial differential equations that represent 
the balance of mass, linear momentum, chemical species,
and energy. These models may include a detailed kinetic 
mechanism for the description of physicochemical phenomena 
involved \cite{lffs1998p577,lffs2000p152,lffs2002p149,lffs2006p353}.
Typically, such reaction mechanisms for mixtures of hydrocarbons 
with air involve tenths of species, hundreds of elementary reactions,
and timescales that vary up to nine orders of magnitude 
\cite{williams1985,elder2011p78,lffs2011p492}.

Concerning the numerical simulation of these models, the challenge
is related to reaction rate of the chemical species, which is nonlinear and 
presents a strong dependence with reaction mechanism dimension. 
Therefore, numerical simulation of a model with a detailed reaction 
mechanism is computationally demanding, which justifies the 
development of techniques to reduce the associated computational cost.

Several techniques to reduce the computational cost of these models
are available in the literature. Essentially, they may be split into two classes. 
The first class includes techniques that seek to reduce the dimension
of reaction mechanisms, such as 
Quasi-Steady State Approximation \cite{frank1940p695},
Rate-Controlled Constrained Equilibrium \cite{keck1990p125},
Computational Singular Perturbation \cite{lam1994p461},
the reduction to Skeleton Mechanisms \cite{law2006},
Intrinsic Low-Dimensional Manifold \cite{maas1992p239}, 
Proper Orthogonal Decomposition \cite{holmes1998}, 
and Invariant Constrained Equilibrium Edge Pre-image Curve
\cite{ren2006p114111}. 
The second class includes techniques that seek efficient 
algorithms to solve complex models, such as 
Look-Up Table \cite{chen1995p505},
Repro-Modeling \cite{turanyi1994p949},
Piece-Wise Reusable Implementation of Solution Mapping
\cite{tonse1999p97}, Artificial Neural Network
\cite{ihme2009p1527}, and \emph{In~Situ} Adaptive Tabulation (ISAT)
\cite{pope1997p41,singer2004p361,liu2005p549,lu2009p361}.

A characterization of one type of transient homogeneous reactor,
using simple and complex thermochemistry, is presented in this
this work. The analysis explores the effects of mixing to residence
time scale ratios as reactor controlling parameter.
The ISAT technique is employed to reduce the computational time 
spent to approximate the solution of the equations that governs 
the evolution of reactive mixture in this transient reactor.
This work also investigates accuracy, performance, and 
memory usage of the present ISAT technique implementation.
The accuracy analysis uses local and global metrics to quantify
the error incurred by ISAT compared with a reference solution
and investigates the statistical seed influence on the results.
The performance is investigated by comparing the computational 
time gain obtained by ISAT with respect to a standard
numerical integration procedure. The memory usage analysis
quantifies the amount of memory used by ISAT implementation.

This work is organized as follows: the transient reactor of interest, 
its governing equation as well the numerical scheme used are 
presented in the second section. The third section presents the 
basic theory of ISAT technique. The fourth section discusses about
the reactor configurations studied and ISAT implementation issues.
Finally, in the fifth section, the main conclusions are summarized
and a path for a future work is suggested.

\section{Transient homogeneous reactor model}

\subsection{The geometry of reactive systems}

In the framework of the transported probability density
function (PDF) models \cite{pope1985p119}, a reactive 
flow may be described by an ensemble of stochastic particles, 
which mimic the behavior of the fluid system.

Consider a transient spatially homogeneous reactive mixture 
evolving adiabatically and at constant pressure in a continuous 
flow reactor. The thermodynamical state of a fluid
particle in this reactor may be completely determined by 
the mass fraction $Y_i$, where  $i=1, \cdots, n_s$, of the $n_s$ 
chemical species and the specific enthalpy $h$, which can be 
lumped in the \emph{composition} vector defined as

\begin{equation}
	\bm{\phi} =
	\transp{\left( h, Y_1, \cdots, Y_{n_s} \right)},
\end{equation}

\noindent
where the superscript $\transp{}$ denotes
the transposition operation. One should note that,
due to the invariance of the system number of atoms,
which ensures the total conservation of the mass, the
components of vector $\bm{\phi}$ are not
linearly independent.

The composition vector of each particle in this flow reactor
evolves according to 

\begin{equation}
	\dod{\bm{\phi}}{t} = 
	- \bm{\Gamma}(\bm{\phi},t) 
	+ \bm{S}(\bm{\phi},t),
	\label{eq_evolution}
\end{equation}

\noindent
where $t$ denotes the time, 
$\bm{\Gamma}$ is the rate of change due to mixing, and 
$\bm{S}$ is the rate of change associated with the chemical reactions.
Integrating  Eq.(\ref{eq_evolution}) from an initial 
time $t_0$ to a time $t$ one obtain the \emph{reaction mapping}

\begin{equation}
	\bm{R}(\bm{\phi}_0,t) =
	\bm{\phi}_0
	- \int_{t_0}^{t} \bm{\Gamma}(\bm{\phi},t')  dt'
	+ \int_{t_0}^{t} \bm{S}(\bm{\phi},t') dt',
	\label{def_RM}
\end{equation}

\noindent
which is the solution of Eq.(\ref{eq_evolution}) 
starting from the initial composition $\bm{\phi}(t_0) = \bm{\phi}_0$.
The reaction mapping corresponds to a trajectory in 
\emph{composition space}, which, for large values of $t$,
tends to the equilibrium composition for the given
enthalpy and mass fractions on $\bm{\phi}_0$.
The composition space is the $(n_s + 1)$--dimensional
Euclidean space where the first direction is associated 
with the enthalpy and the remaining $n_s$ are related to the 
chemical species.

\subsection{Pairwise mixing stirred reactor}
\label{pmsr}

The classical \emph{Partially Stirred Reactor} (PaSR), 
used in \cite{correa1993p41,lffs2002p164}, describes $\bm{\Gamma}$
by the interaction by exchange with the mean
(IEM) micromixing model \cite{fox2003}, but, 
for the purpose of testing a thermochemistry reduction 
technique, it is desirable to employ a mixing model that leads 
to a composition region accessed during the solution process
which is ``wider'' than that provided by the IEM model.
A modified version of PaSR model called  
\emph{Pairwise Mixing Stirred Reactor} (PMSR),
\cite{pope1997p41}, is designed to yield 
a much larger accessed region, and, hence, should 
provide a stringent test on the ability of ISAT technique 
to yield a reduction in computational time.

In the PMSR model the reactor consists of an even number 
$n_{p}$ of particles, initially arranged in pairs $(j_1,j_2)$
such that the particles $(1,2)$, $(3,4)$, $\cdots$,
$(n_{p}-1,n_{p})$ are partners. Given a time step, 
$\Delta t$, for each discrete time $k\Delta t$, 
where $k$ is an integer, the model is characterized by
three types of events: \emph{inflow}, 
\emph{outflow}, and \emph{pairing}.
The inflow and outflow events consist of randomly selecting 
\mbox{$n_{in} = \ceil{ \frac{1}{2} n_{p} (\Delta t / \tau_r) }$}
pairs of particles, being $\tau_r$ the residence time within the
reactor, and exchanging their thermodynamical properties 
by the properties of a prescribed inflow. 
The pairing event consists of randomly selecting for pairing
a number of pairs of particles, different from the inflow particles, equal to
\mbox{$n_{pair} = \ceil{ \frac{1}{2} n_{p} (\Delta t / \tau_p) }$},
being $\tau_p$ the pairwise time. Then the chosen particles 
(inflow/outflow and paring) are randomly shuffled. Between these 
discrete times, the pairs of particles $(j_1,j_2)$ evolve according to

\begin{equation}
	\dod{\bm{\phi}}{t}^{(j_1)}  =  
	- \frac{\bm{\phi}^{(j_1)} - \bm{\phi}^{(j_2)}}{\tau_m}
	+ \bm{S}(\bm{\phi}^{(j_1)},t),
	\label{pmsr_eq1}
\end{equation}

\begin{equation}
	\dod{\bm{\phi}}{t}^{(j_2)}  =  
	- \frac{\bm{\phi}^{(j_2)} - \bm{\phi}^{(j_1)}}{\tau_m}
	+ \bm{S}(\bm{\phi}^{(j_2)},t).
	\label{pmsr_eq2}
\end{equation}

\noindent
where $\tau_m$ is the mixing time.

\subsection{Numerical Integration}
\label{num_int}

An operator splitting technique
\cite{yang1998p16} is employed
to numerically integrate Eq.(\ref{eq_evolution}).
The overall process of integration via operator
splitting technique can be represented as

\begin{equation}
    \bm{\phi}(t) \stackrel{mixing}{\longrightarrow} 
    \bm{\phi}_{mix}(t + \Delta t) \stackrel{reaction}{\longrightarrow}
    \bm{\phi}(t + \Delta t),
    \label{op_split}
\end{equation}

\noindent
where given an initial composition $\bm{\phi}_0$ 
and a time step $\Delta t$, the first fractional step
integrates the pure mixing system, 

\begin{equation}
	\dod{\bm{\phi}}{t} = - \bm{\Gamma}(\bm{\phi},t) ,
	\label{eq_mixing}
\end{equation}

\noindent
to obtain $\bm{\phi}^{(j)}_{mix}(t + \Delta t)$. Then, 
the pure chemical reaction system, 

\begin{equation}
	\dod{\bm{\phi}}{t} = \bm{S}(\bm{\phi},t).
	\label{eq_reaction}
\end{equation}

\noindent
is integrated from an initial composition 
$\bm{\phi}_{mix}(t + \Delta t)$
over a time step $\Delta t$ and gives
$\bm{\phi}(t + \Delta t)$.

The operator splitting technique allows to solve each term in the 
evolution equation, Eq.(\ref{eq_evolution}), separately, 
using specific efficient numerical methods to treat the particular 
features inherent to the physical phenomenon modeled by 
each term \cite{fox2003}.

\section{In situ adaptive tabulation}

\subsection{Linearized reaction mapping}

Consider a composition $\bm{\phi}$
and an \emph{initial composition} $\bm{\phi}_0$,
so that Taylor expansion of the reaction 
mapping of $\bm{\phi}$ around $\bm{\phi}_0$ is

\begin{equation}
    \bm{R}(\bm{\phi},t) = \bm{R}(\bm{\phi}_0,t) +
    \bm{A}(\bm{\phi}_0,t) \delta \bm{\phi} +
    \order{\norm{\delta \bm{\phi}}^2},
    \label{reac_map_series}
\end{equation}

\noindent
where $\delta \bm{\phi} = \bm{\phi} - \bm{\phi}_0$,
$\bm{A}(\bm{\phi}_0,t)$ is a $n_{\phi} \times n_{\phi}$ matrix,
called \emph{mapping gradient matrix}, with components given by

\begin{equation}
    A_{ij}(\bm{\phi}_0,t) = 
    \dpd{R_i}{\phi_{0_{j}}}(\bm{\phi}_0,t),
\end{equation}

\noindent
$\norm{\cdot}$ denotes the Euclidean norm of a vector,
and $\order{\norm{\delta \bm{\phi}}^2}$ denotes terms 
that have order $\norm{\delta \bm{\phi}}^2$.

The \emph{linear approximation} $\bm{R}^l(\bm{\phi},t)$,
obtained by neglecting the high order terms of 
Eq.(\ref{reac_map_series}), is second-order accurate at 
a connected region of composition space centered at 
$\bm{\phi}_0$.  The shape of this region is unknown before 
the calculations, but ISAT algorithm approximates this 
region by a hyper-ellipsoid, as will be shown in section
\ref{ell_of_accur}. The \emph{local error} of this 
linear approximation is defined as the Euclidean norm of the 
difference between the reaction mapping
at $\bm{\phi}$ and the linear approximation
for it around $\bm{\phi}_0$,

\begin{equation}
	\varepsilon =
	\norm{ \bm{R}(\bm{\phi},t)-\bm{R}^l(\bm{\phi},t) }.
	\label{local_error}
\end{equation}

\subsection{Ellipsoid of accuracy}
\label{ell_of_accur}

The accuracy of the linear approximation
at $\bm{\phi}_0$ is controlled only if the 
local error is smaller than a
positive \emph{error tolerance} $\epstol$,
which is heuristically chosen. 
The \emph{region of accuracy}
is defined as the connected region of the
composition space centered at $\bm{\phi}_0$ where local error is not 
greater than $\epstol$. As shown in \cite{pope1997p41},
this region is approximated by a hyper-ellipsoid  centered 
at $\bm{\phi}_0$ which is dubbed \emph{ellipsoid of accuracy}
(EOA), and is mathematically represented by the following inequation

\begin{equation}
	\delta \transp{\bm{\phi}} \bm{L} 
	\transp{\bm{L}} \delta \bm{\phi}
	\leq \epstol^2,
	\label{chol_eoa_eq}
\end{equation}

\noindent
where the EOA Cholesky matrix, denoted by $\bm{L}$,
is lower triangular \cite{golub1996}.

The adaptive step of ISAT algorithm
involves the solution of the following
geometric problem: given a hyper-ellipsoid
centered at $\bm{\phi}_0$ and a 
\emph{query composition}, $\bm{\phi}_q$, 
outside it, determine a new hyper-ellipsoid 
of minimum hyper-volume, centered at $\bm{\phi}_0$, 
which encloses both the original hyper-ellipsoid and 
the point $\bm{\phi}_q$. The solution of this problem
is presented by \cite{pope2008} and is
not shown here for sake of brevity.

\subsection{Tabulation procedure}

Initially ISAT algorithm receives the 
time step $\Delta t$ and the tolerance $\epstol$.
Then, in every time step, ISAT algorithm 
receives a query composition $\bm{\phi}_q$ and returns 
an approximation for the corresponding reaction mapping 
$\bm{R}(\bm{\phi}_q,t)$. This approximation is obtained 
via numerical integration of Eq.(\ref{eq_reaction}) or 
by the linear approximation $\bm{R}^l(\bm{\phi}_q,t)$.

During the reactive flow calculation, the computed 
values are sequentially stored in a 
table for future use. This process is known as 
\emph{in~situ} tabulation. The ISAT table, which is created by 
the tabulation process, includes the initial composition
$\bm{\phi}_0$, the reaction mapping
$\bm{R}(\bm{\phi}_0,t)$, and the mapping gradient 
matrix $\bm{A}(\bm{\phi}_0,t)$. Using these 
elements it is possible to construct the 
linear approximation.
As the calculation proceeds, a new query composition, 
$\bm{\phi}_q$, is received by ISAT and the table is 
transversed until a $\bm{\phi}_0$ 
is found that is close to $\bm{\phi}_q$. 
Depending on the accuracy, the linear 
approximation around $\bm{\phi}_0$ is returned or 
the reaction mapping of $\bm{\phi}_q$ is obtained by 
direct integration of Eq.(\ref{eq_reaction}).

The ISAT table is a binary search tree, since this data 
structure allows for searching an information
in $\order{\log_2{n_{tab}}}$ operations, where $n_{tab}$ is the total 
number entries in the tree, if the tree is balanced \cite{knuth_v3}.
The binary search tree is basically formed by two types of 
elements, nodes, and leaves. Each leaf of the tree stores the 
following data:

\begin{itemize}
	\item $\bm{\phi}_0$: initial composition;
	\item $\bm{R}(\bm{\phi}_0,t)$: reaction mapping at $\bm{\phi}_0$;
	\item $\bm{A}(\bm{\phi}_0,t)$: mapping gradient matrix at $\bm{\phi}_0$;
	\item $\bm{L}$: EOA Cholesky matrix.
\end{itemize}

\begin{figure} [ht!]
	\centering
	\includegraphics[scale=0.8]{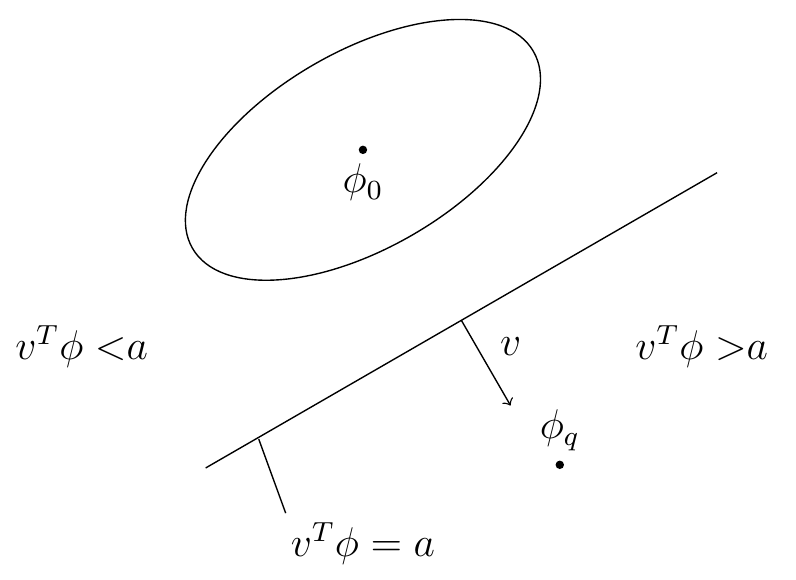}
	\caption{Cutting plane in relation to EOA position.}
	\label{cutting}
\end{figure}

\noindent
Each node of the binary search tree has an associated
\emph{cutting plane}. This plane is defined by a
\emph{normal vector}

\begin{equation}
	\bm{v} =
	\bm{\phi}_q - \bm{\phi}_0,
\end{equation}

\noindent
and a scalar

\begin{equation}
	a =
	\transp{\bm{v}} 
	\left( \frac{\bm{\phi}_q + \bm{\phi}_0}{2} \right),
\end{equation}

\noindent
such that all compositions $\bm{\phi}$ with 
$\bm{v}^T\bm{\phi} > a$ are located to the
``right" of the cutting plane, as sketched in 
Figure~\ref{cutting}. The cutting plane construction 
defines a search criterion in the binary search tree.

If during the calculation a query point $\bm{\phi}_q$
is encountered within the region of accuracy,
i.e., $\varepsilon \leq \epstol$, 
but outside the estimate of EOA, then the EOA growth proceeds 
as detailed in \cite{pope2008}.
The first three items stored in the binary search tree leaf 
$\left[  \bm{\phi}_0,~\bm{R}(\bm{\phi}_0,t)~\mbox{and}~\bm{A}(\bm{\phi}_0,t) \right]$
are computed once, whereas $\bm{L}$ changes whenever the EOA is grown.

Once a query composition $\bm{\phi}_q$ is received by ISAT table,
the binary search tree is initialized as a single leaf 
and the exact value of the reaction mapping is returned.

\begin{figure} [h!]
	\centering
	\includegraphics[scale=0.8]{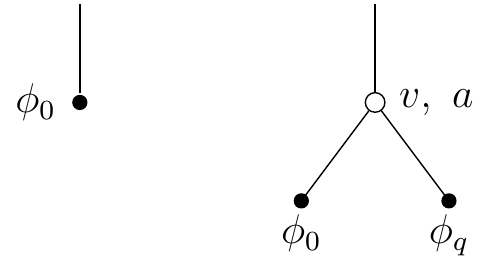}
	\caption{Binary search tree before and after the addition of a new node.}
	\label{nodes}
\end{figure}

The subsequent steps are:

\begin{enumerate}

\item Given a query composition the tree is
	transversed until a leaf ($\bm{\phi}_0$) is found.

\item Equation (\ref{chol_eoa_eq}) is used to determine
	if $\bm{\phi}_q$ is inside EOA or not.

\item If $\bm{\phi}_q$ is inside EOA, the reaction mapping
	is given by the linear approximation.
	This is the first of four outcomes, called \emph{retrieve}.

\item If $\bm{\phi}_q$ is outside EOA, direct integration is used
	to compute the reaction mapping, and
	the local error is measured by Eq.(\ref{local_error}).

\item If the local error is smaller than tolerance, $\epstol$,
	the EOA is grown according to the procedure presented
	in \cite{pope2008} and the reaction mapping
	is returned. This outcome is called \emph{growth}.

\item If local error is greater than the tolerance 
    $\epstol$ and the maximum number
	of entries in the binary search tree is not reached,
	a new record is stored in the binary search tree based 
	on $\bm{\phi}_q$ and the reaction mapping
	is returned. The original leaf is replaced by a 
	node with the left leaf representing the old composition 
	$\bm{\phi}_0$ and the right leaf the new one $\bm{\phi}_q$
	as shown in Figure~\ref{nodes}. This outcome is an 
	\emph{addition}.

\item If the local error is greater than the tolerance 
	$\epstol$ and the maximum number
	of entries in the binary search tree is reacted,
	the reaction mapping is returned.
	This outcome is called \emph{direct evaluation}.
	
\end{enumerate}

\section{Results and discussion}

This section presents the results of numerical simulations performed
in order to assess characteristics of PMSR reactor. This study investigates 
three configurations of a PMSR reactor, two mixtures of $CO/O_{2}$, and 
one mixture of $CH_4$/air, using different time scales to define 
their behavior.
Accuracy, performance, and memory usage of ISAT technique implementation are
also investigated. For this purpose, the approximation for solution of the
model equation, Eq.(\ref{eq_evolution}), obtained by ISAT technique
are compared with a reference solution from direct integration (DI) procedure, 
described in section~\ref{num_int}.

\subsection{$CO/O_{2}$ mixture}

The first two cases studied consider a 1024 particles PMSR,
initially filled with a fuel-lean mixture of $CO/O_{2}$ at 
$2948.5~\unit{K}$ and $1~\unit{atm}$. 
At every time step, a fuel-lean mixture of $CO/O_{2}$ enters 
the reactor at $T_{in} = 300~\unit{K}$ and $p_{in} = 1~\unit{atm}$. 
Both fuel-leam mixtures have equivalence ratio equal to $0.7$.
The constant pressure and enthalpy equilibrium state associated 
to the inflow mixture is reached at $T_{eq} = 2948.5~\unit{K}$. 
The reaction mechanism used to describe $CO$ with $O_2$ kinetics 
involves 4 species and 3 reactions, \cite{gardiner2000}. 
For the first case a time scales configuration with
$\tau_{m}/\tau_{r} = \tau_{p}/\tau_{r} = 1/2$ is used, 
so that the pairwise/mixing time scales are of the same order 
of magnitude as the residence time. In this situation, 
partially stirred reactor (PaSR) conditions result within
the reactor. For the second case, time scales configuration 
adopted is $\tau_{m}/\tau_{r} = \tau_{p}/\tau_{r} = 1/10$,
so that the pairwise/mixing time scales are small when compared to
the residence time. Thus, the reactor should behave almost as a 
perfect stirred reactor (PSR), where the processes of mixing and
pairing occur instantaneously. These cases use a binary search 
tree with a maximum of $50,000$ entries, $\Delta t = 10~\unit{\mu s}$,
and $\epstol=10^{-3}$.

\subsection{$CH_4$/air mixture}

The third case studied consists of a PMSR, with 1024 particles,
initially filled with the combustion products of a stoichiometric mixture 
of $CH_4$/air at $2100~\unit{K}$ and $1~\unit{atm}$.
At every time step, a stoichiometric mixture 
of $CH_4$/air enters the reactor at $T_{in} = 300~\unit{K}$
and $p_{in} = 1~\unit{atm}$. The constant pressure and 
enthalpy equilibrium state associated to this mixture is
reached at $T_{eq} = 2225.5~\unit{K}$.
The reaction of $CH_{4}$ with air kinetics is described 
by GRI mechanism version~3.0 \cite{grimech},
with 53 species and 325 reactions. 
For this PMSR one have
$\tau_{m}/\tau_{r} = \tau_{p}/\tau_{r} = 1/4$,
so that it is expected to behave like a PaSR.
This simulation uses a binary search tree with a maximum 
of 60,000 entries, $\Delta t = 0.1~\unit{ms}$ and $\epstol=10^{-3}$.

\subsection{Analysis of reactor behavior}
\label{analysis_pmsr_behav}

\begin{figure*}[ht]
    \centering
    \subfigure[Evolution of $\dless{\mean{T}}$ for case~1.]
    {\includegraphics[scale=0.35]{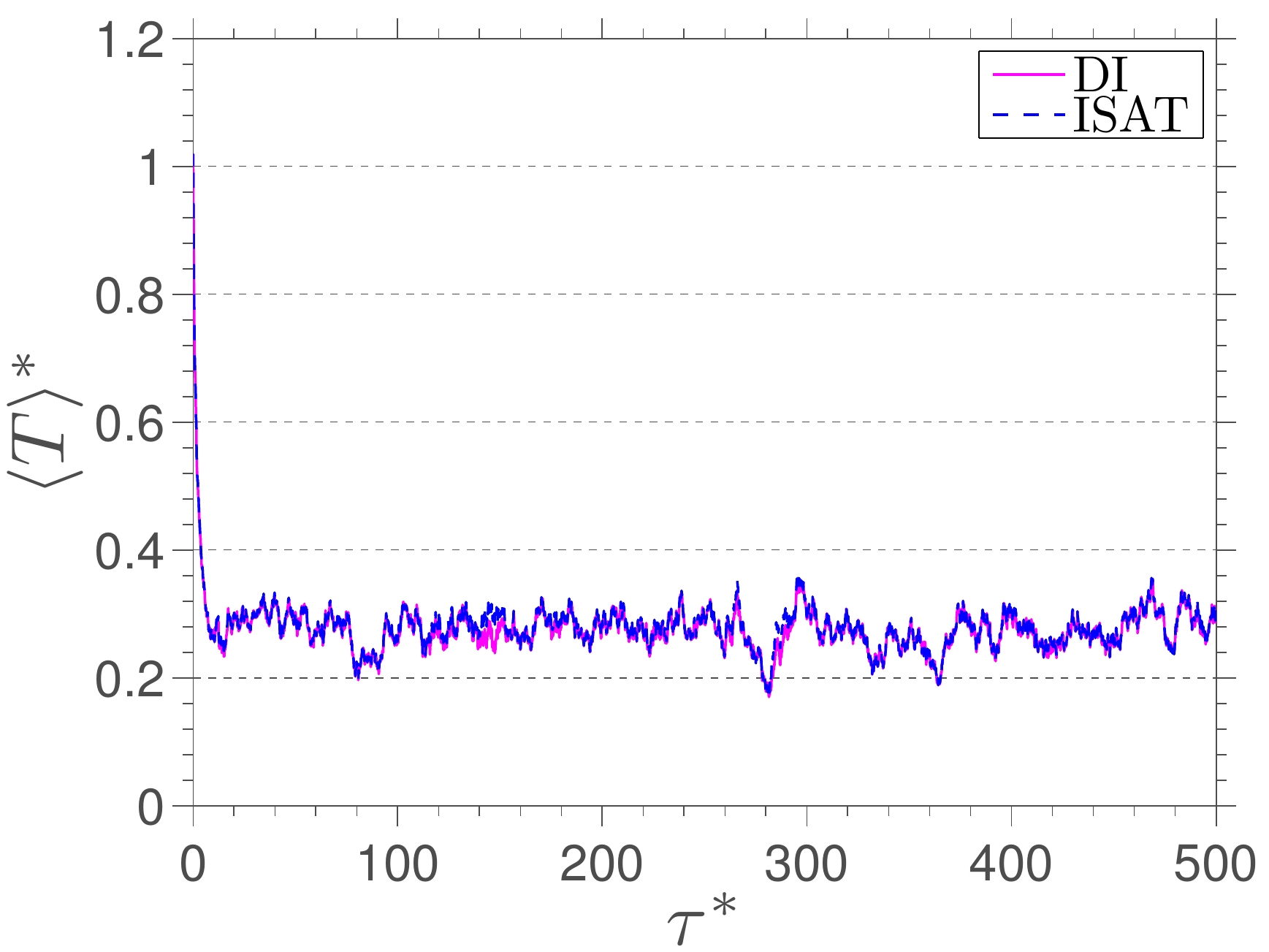}}
    \subfigure[Evolution of $\dless{\mean{T}}$ for case~2.]
    {\includegraphics[scale=0.35]{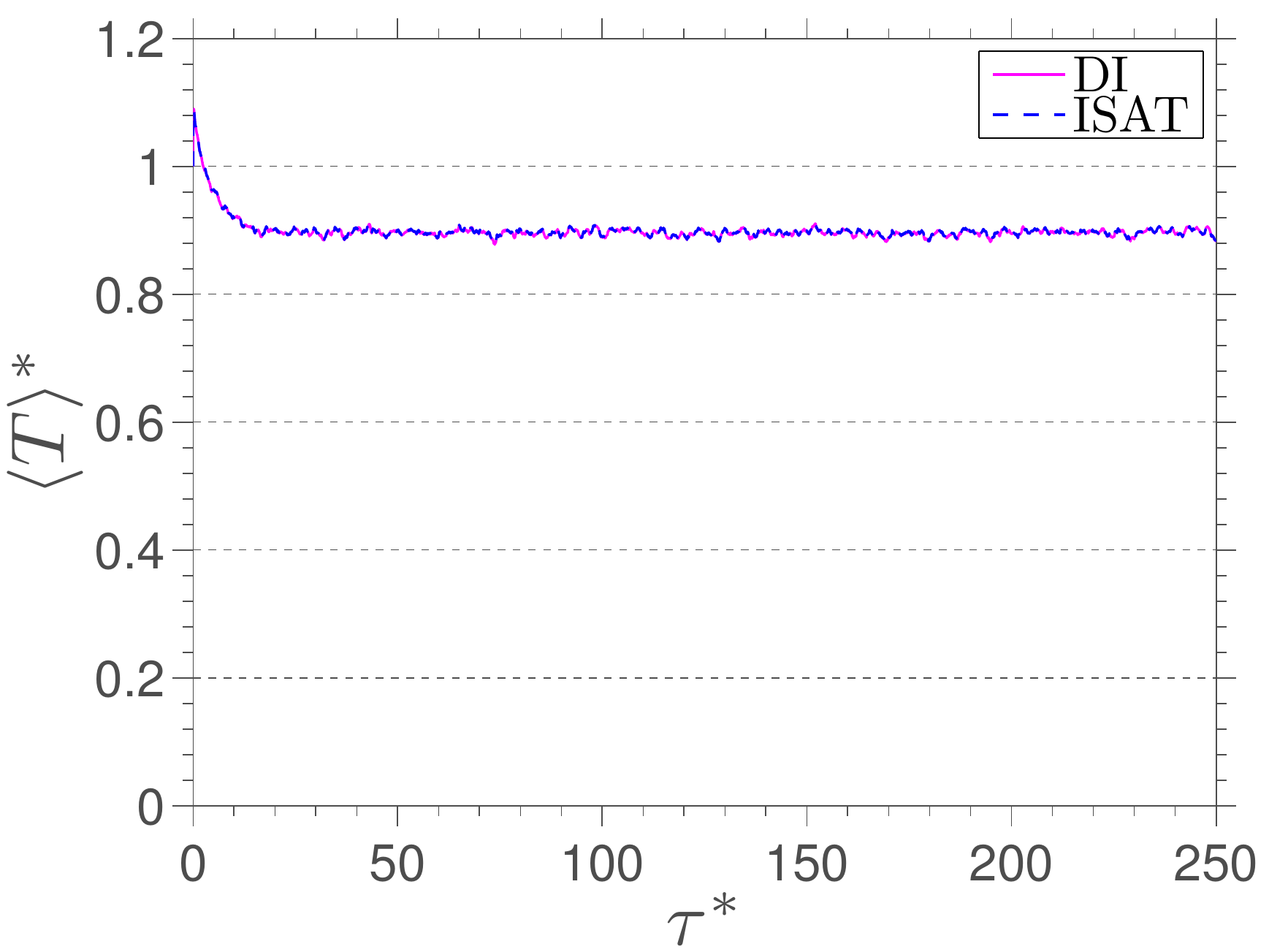}}
    \caption{Comparison between DI (---) and ISAT (- - -) calculations 
    of the ensemble average of the reduced temperature.}
    \label{cases12_T_mean_vs_t}
\end{figure*}

\begin{figure*}[ht!]
    \centering
    \subfigure[Evolution of $\dless{\var{T}}$ for case~1.]
    {\includegraphics[scale=0.35]{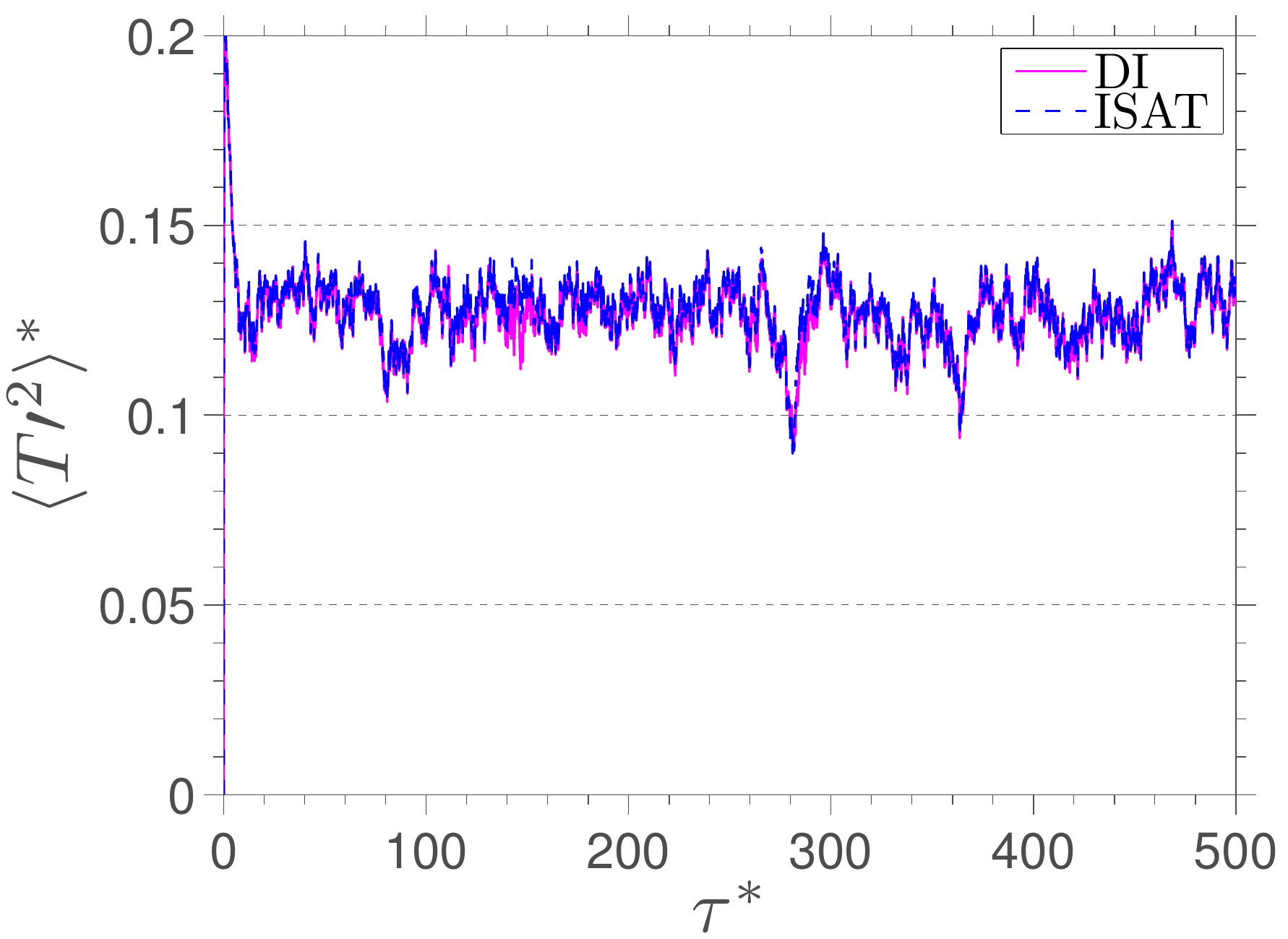}}
    \subfigure[Evolution of $\dless{\var{T}}$ for case~2.]
    {\includegraphics[scale=0.35]{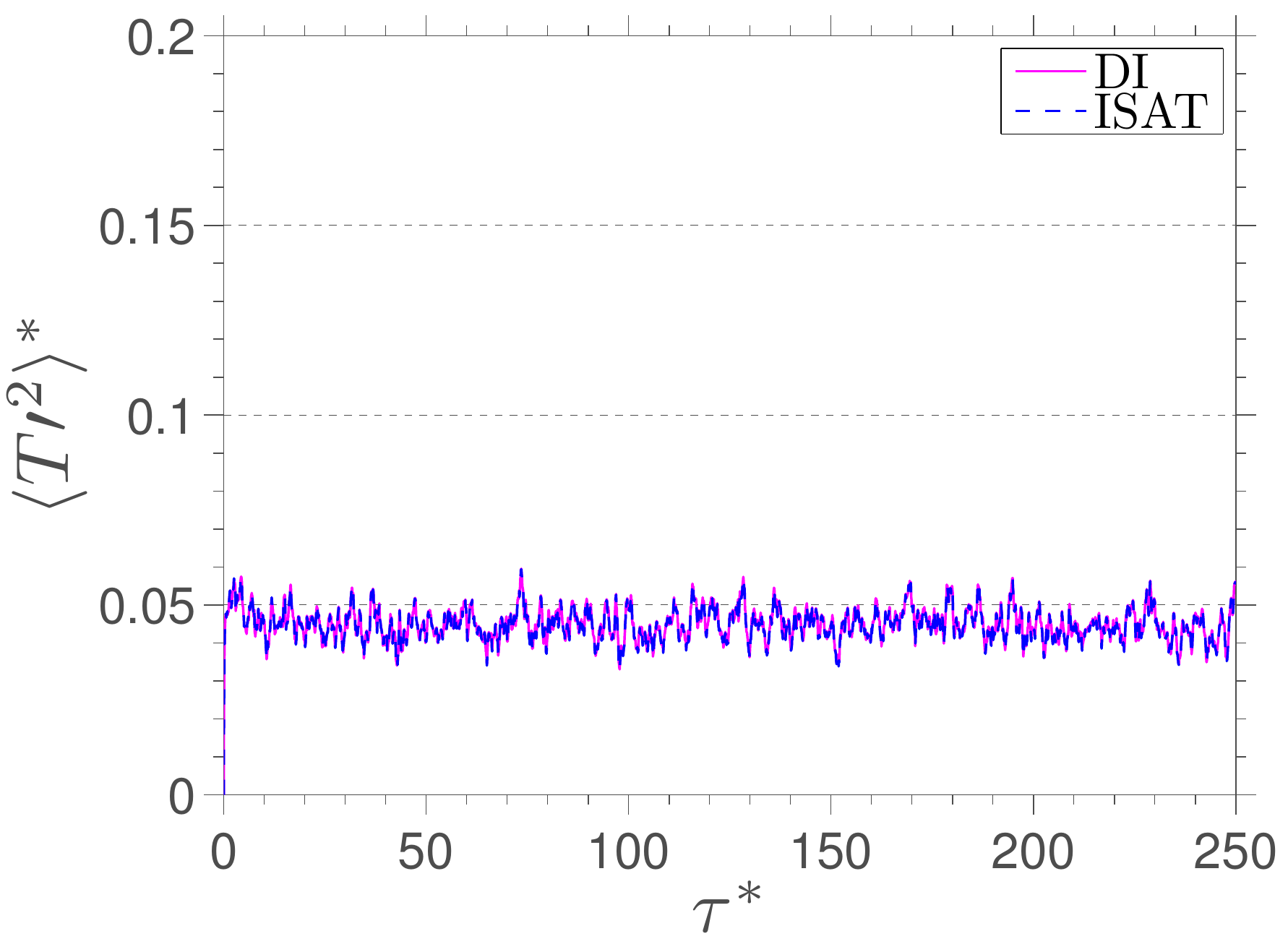}}
    \caption{Comparison between DI (---) and ISAT (- - -) calculations
    of the ensemble variance of the reduced temperature.}
    \label{cases12_T_var_vs_t}
\end{figure*}

Figures~\ref{cases12_T_mean_vs_t}~and~\ref{cases12_T_var_vs_t} 
show the comparison between DI and ISAT calculations, 
as function of the dimensionless time 
$\dless{\tau} = t / \tau_r$, for the ensemble average
$\dless{\mean{T}}$ and the ensemble variance $\dless{\var{T}}$
of the reduced temperature in the cases~1~and~2, where the reduced 
temperature is defined as

\begin{equation}
    \dless{T} =
    \frac{T - T_{in}}{T_{eq} - T_{in}},
    \label{def_dless_T}
\end{equation}

\noindent
and the ensemble average and the ensemble variance operators 
are respectively defined as

\begin{equation}
	\mean{\psi} =
	\frac{1}{n_p} \sum_{j=1}^{n_p} \psi^{(j)}
	\qquad \mbox{and} \qquad
	\var{\psi} = \mean{\psi^2} - \mean{\psi}^2,
	\label{def_mean_var}
\end{equation}

\noindent
being $\psi$ a generic property of the reactive system.

For case~1 results, which span over a range of $500$
residence times, one can observe an excellent qualitative 
agreement for $\dless{\mean{T}}$ and $\dless{\var{T}}$.
The ensemble average value rapidly drops from the initial value 
to reach the statistically steady-state regime around 
$\dless{\mean{T}} = 0.3$.
The ensemble variance rapidly grows, then decreases until 
it reaches the statistically steady-state value around 
$\dless{\var{T}} = 0.13$. The analysis of the two figures 
shows that the statistically steady-state regime is 
reached after $\dless{\tau} = 10$.

In case~2, where a range of $250$ residence times have been
computed, one can also observe an excellent qualitative 
agreement for $\dless{\mean{T}}$ and $\dless{\var{T}}$.
Again, the overall history of the PMSR is the same for DI and ISAT. 
Similar results, not shown here for sake of brevity, have been obtained 
for the other thermochemical properties of the reactors.

\begin{figure*}[ht!]
    \centering
    \subfigure[Histogram of $\dless{T}$ for case~1.]
    {\includegraphics[scale=0.35]{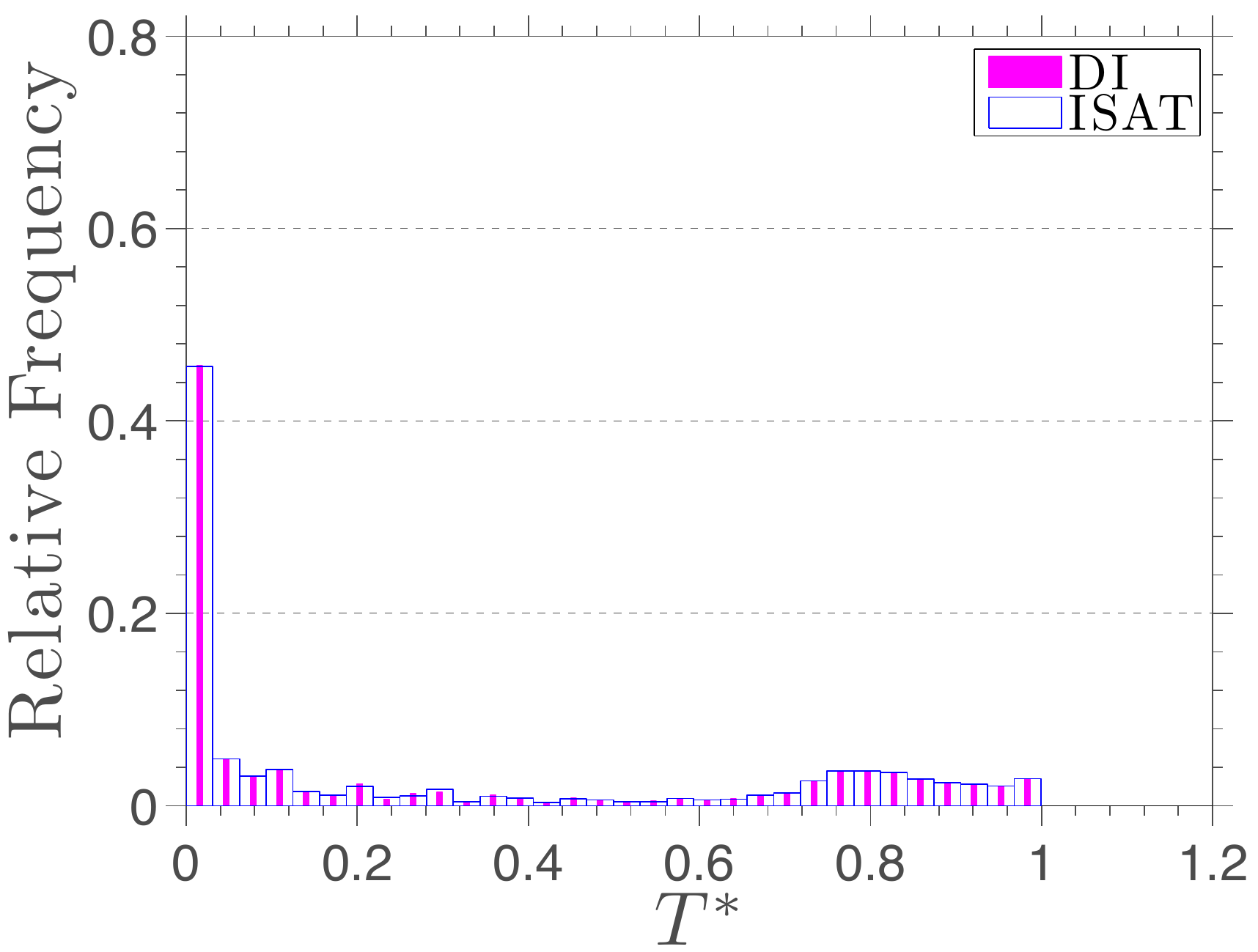}}
    \subfigure[Histogram of $\dless{T}$ for case~2.]
    {\includegraphics[scale=0.35]{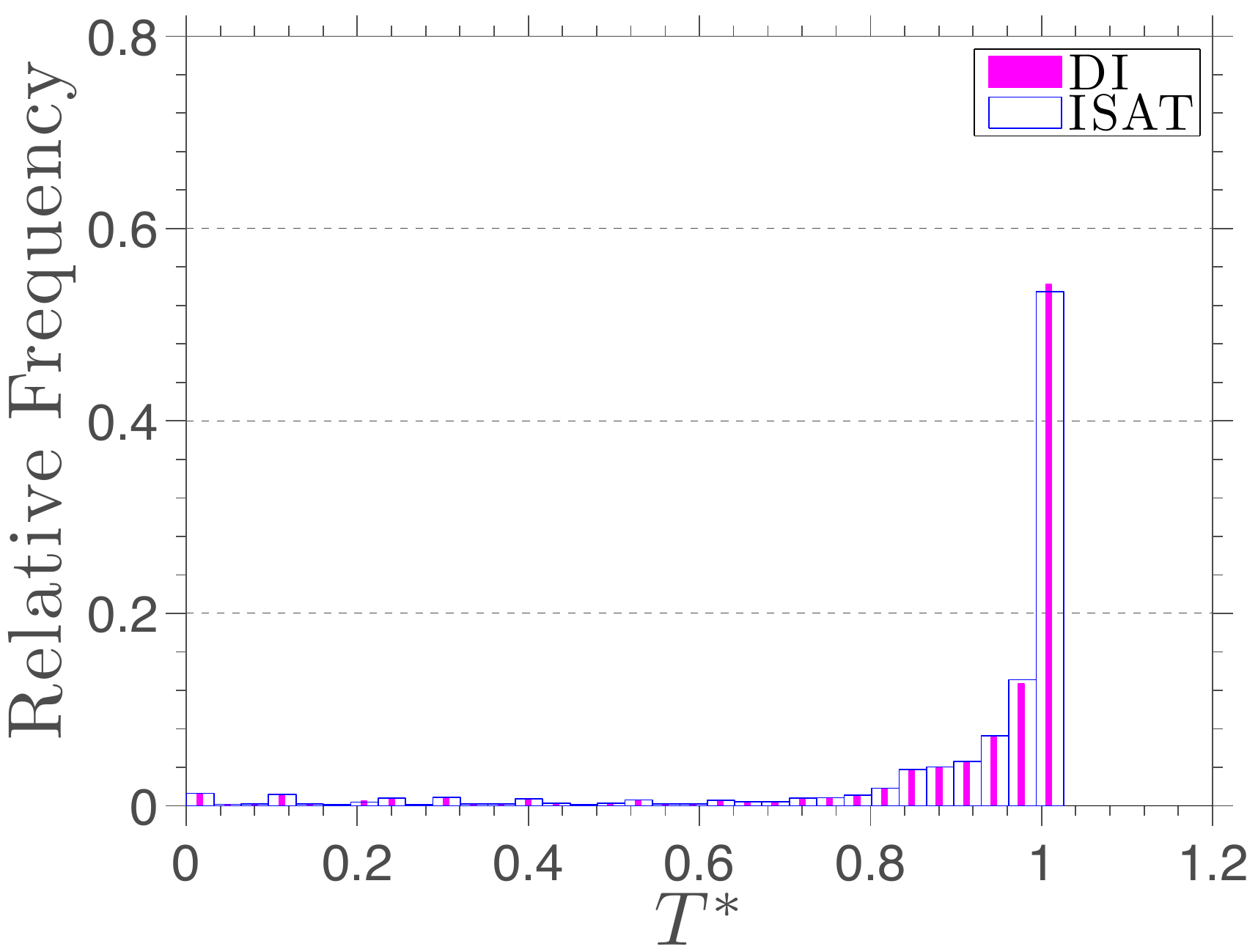}}
    \caption{Comparison between DI and ISAT calculations of the 
    mean histograms (averaged over the last $50$ residence times)
    of the reduced temperature.}
    \label{cases12_pdf_T}
\end{figure*}

\begin{figure*}[ht!]
    \centering
    \subfigure[Histogram of $Y_{O}$ for case~1.]
    {\includegraphics[scale=0.35]{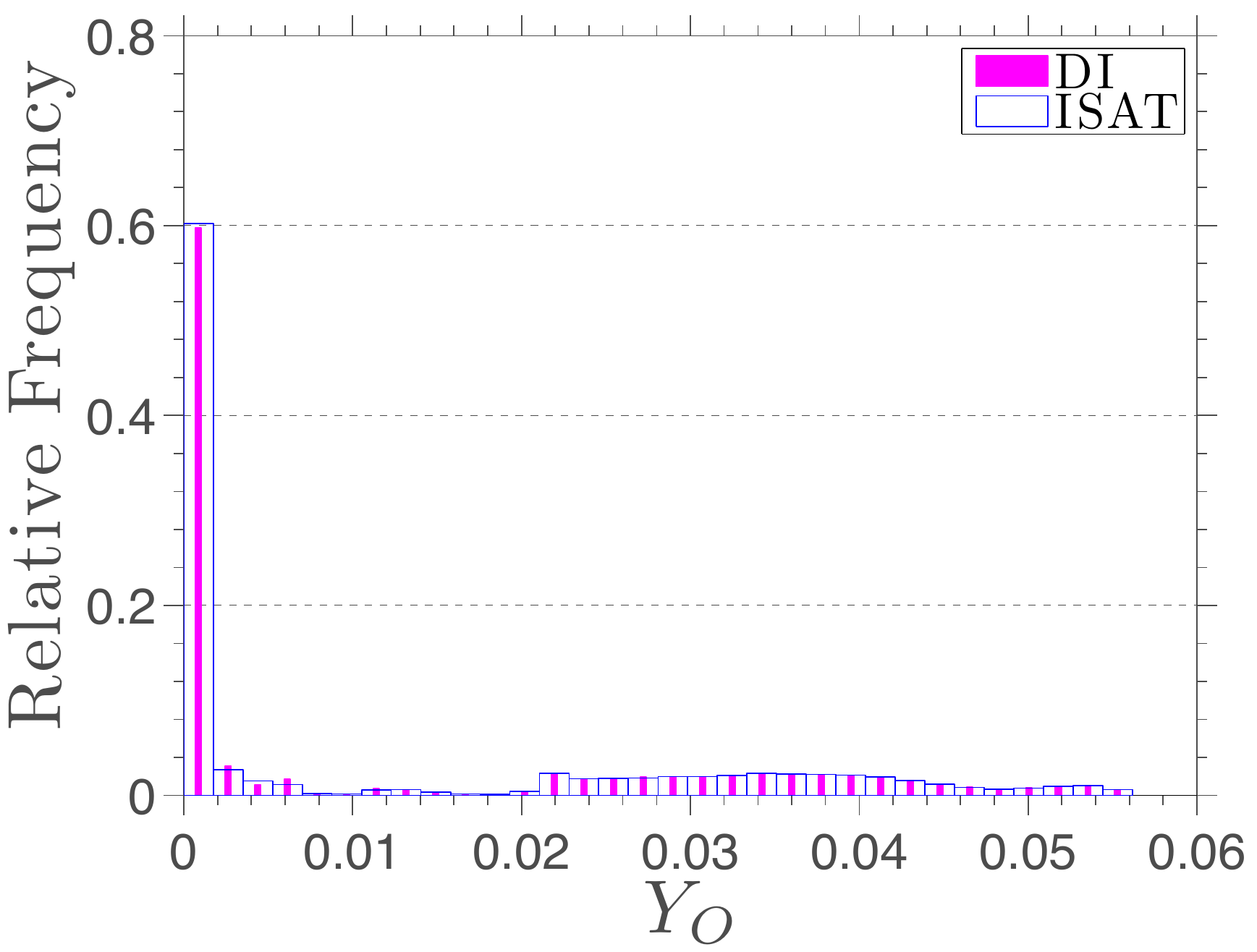}}
    \subfigure[Histogram of $Y_{O}$ for case~2.]
    {\includegraphics[scale=0.35]{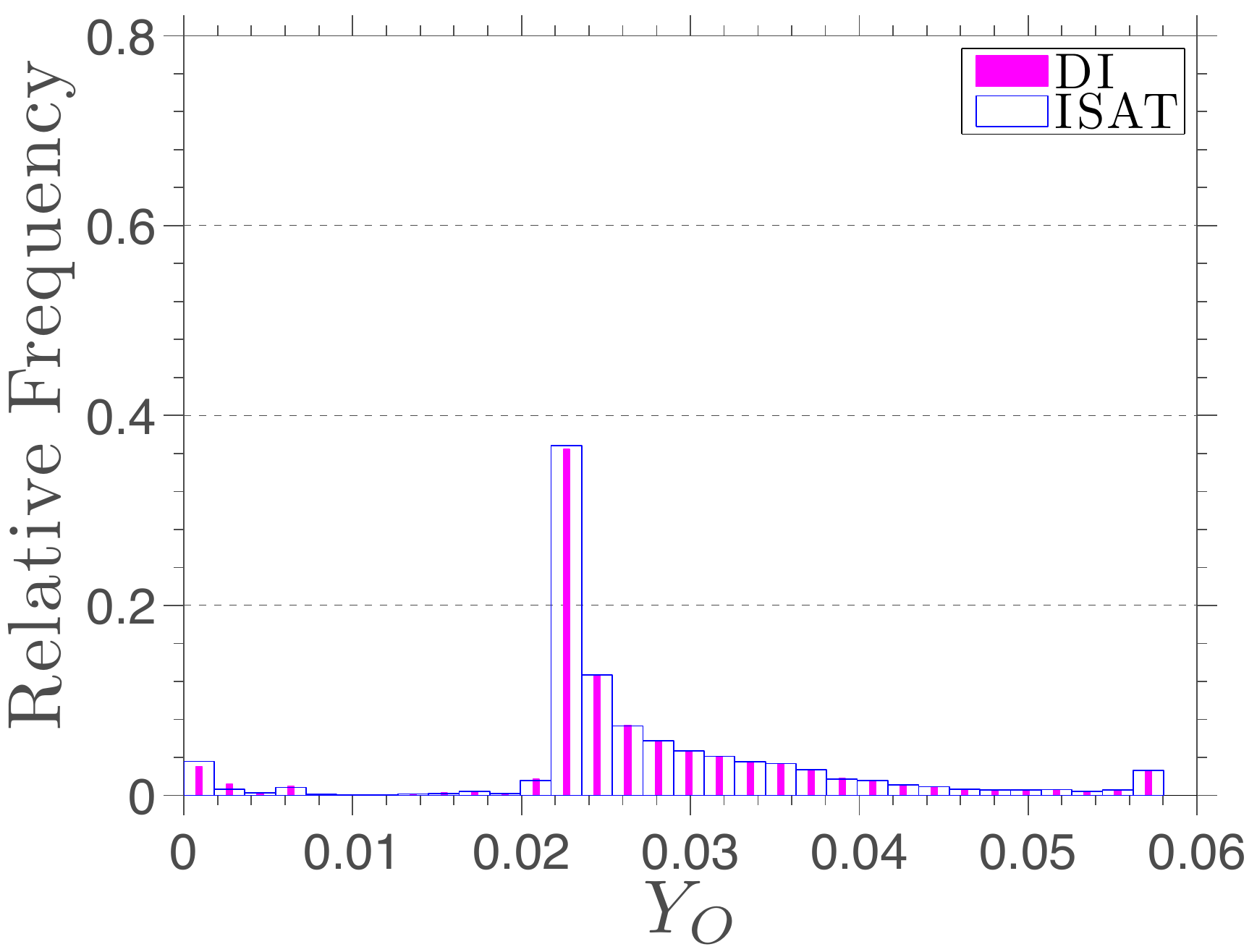}}
    \caption{Comparison between DI and ISAT calculations of the 
    mean histograms (averaged over the last $50$ residence times)
    of the $O$ mass fraction.}
    \label{cases12_pdf_O}
\end{figure*}

The difference among both cases is due to the behavior
of each reactor at the statistically steady state regime. 
Indeed, the case~2 reactor behavior is governed by a 
competition between the chemical and residence times mostly.
Therefore, the thermodynamical properties steady-state
PDF is concentrated over a smaller range
than in case~1, where the mixing and pairing time scales are
large when compared to the residence time.
This behavior is illustrated in 
Figures~\ref{cases12_pdf_T}~and~\ref{cases12_pdf_O},
which present the comparison between DI and ISAT computations 
of mean histograms, averaged over the last $50$ residence times,
of $\dless{T}$ and $Y_O$ for cases~1~and~2. 
These figures underscore the influence of PMSR controlling parameters,
i.e., the time scales ratios $\tau_m / \tau_r$ and $\tau_p / \tau_r$, 
on the thermochemical conditions prevailing within each reactor. 
Indeed, the temperature within the case~2 reactor is such that almost 
only burned gases are found. On the other hand, case~1 reactor is 
characterized by a bimodal temperature distribution with a large 
probability of finding $\dless{T} = 0.04$ and a broader temperature 
distribution leaning toward the burned gases. Such a distribution, 
as it could be expected, is reflected on $Y_{O}$ histogram, which also 
exhibits a bimodal distribution.

\begin{figure*}[ht!]
    \centering
    \subfigure[Evolution of $\dless{\mean{T}}$ for case~3.]
    {\includegraphics[scale=0.35]{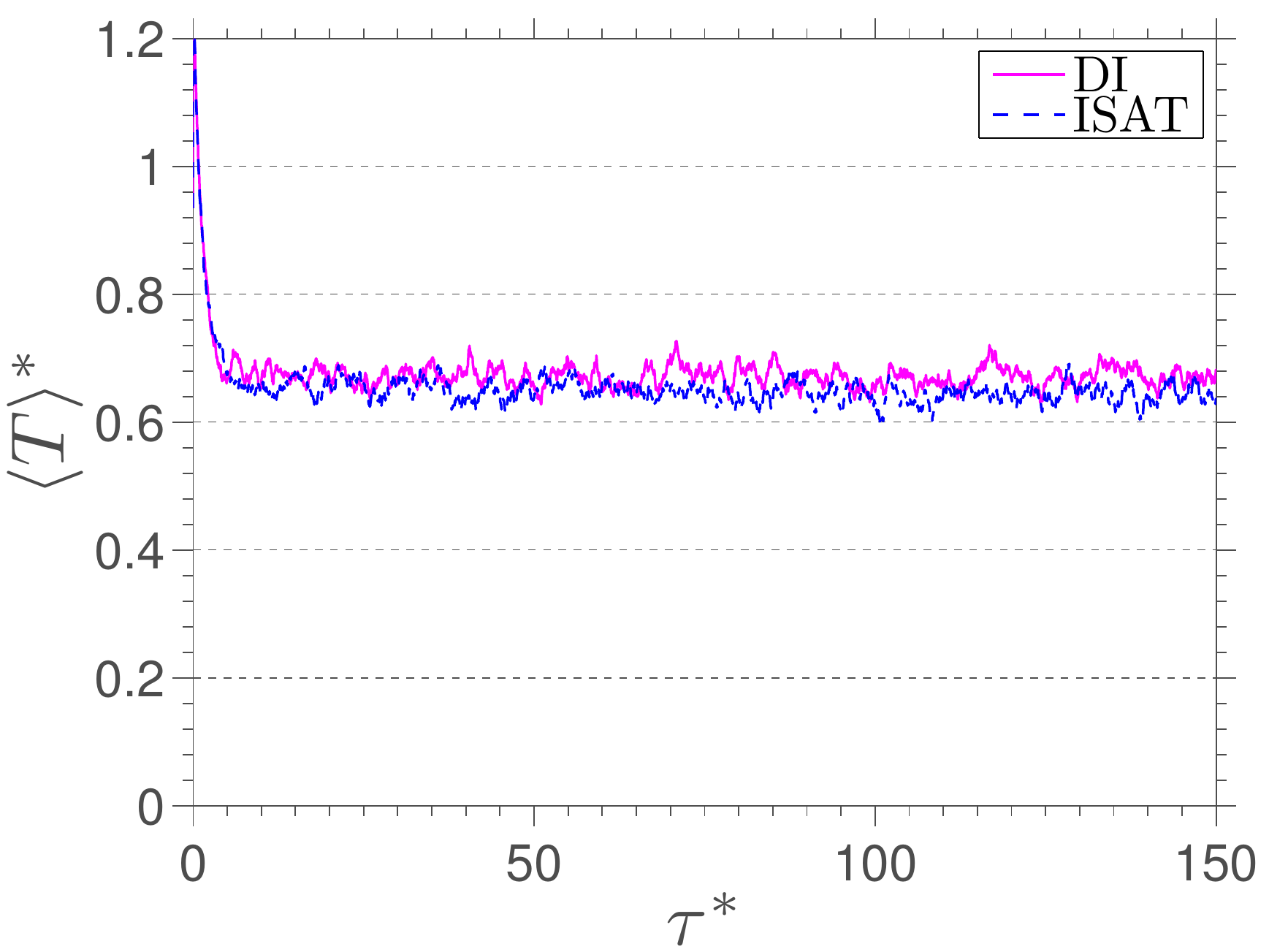}}
    \subfigure[Evolution of $\mean{Y_{OH}}$ for case~3.]
    {\includegraphics[scale=0.35]{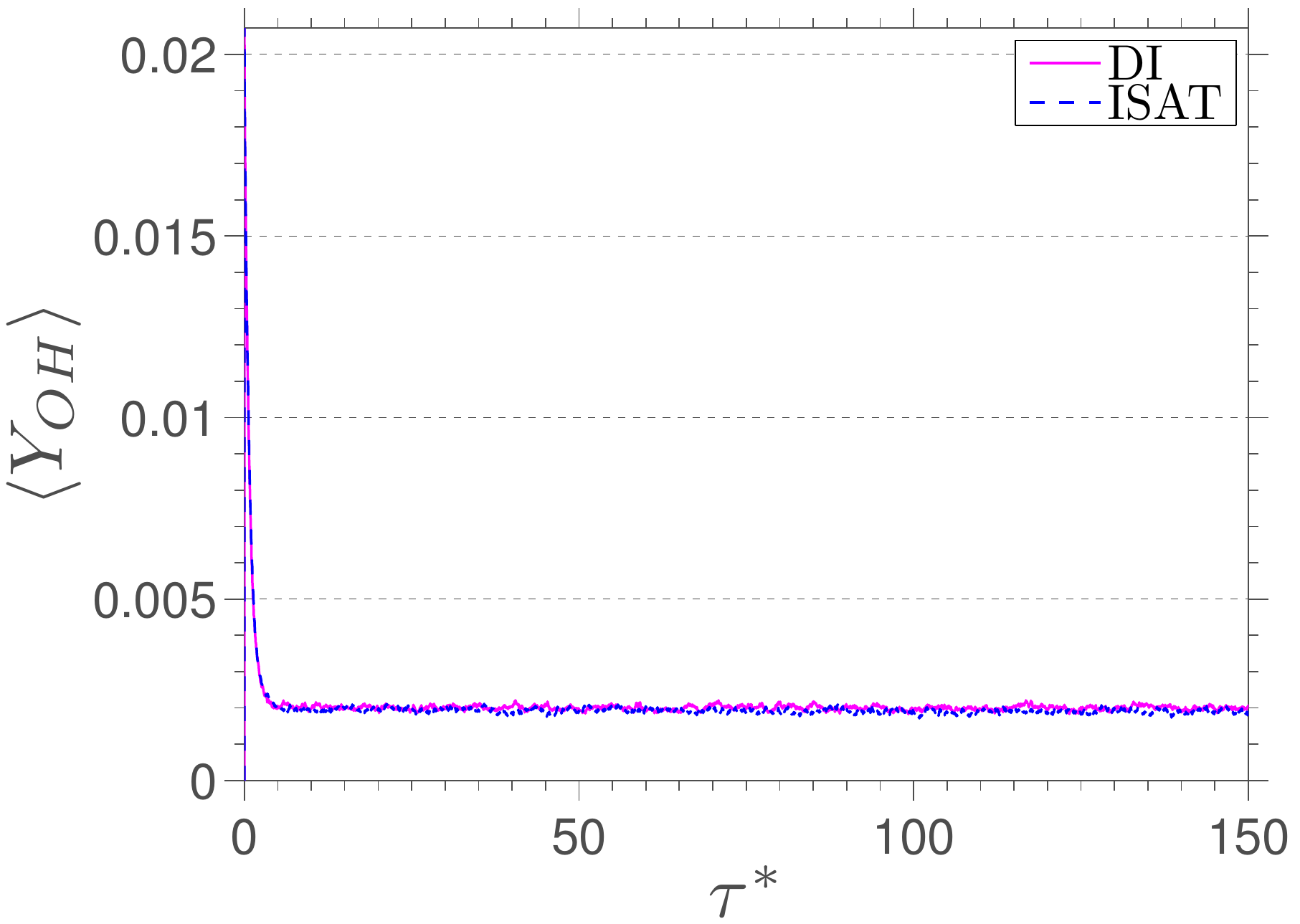}}
    \caption{Comparison between DI (---) and ISAT (- - -) calculations 
    of the ensemble average of the reduced temperature and $OH$ mass fraction.}
    \label{case3_T_OH_mean_vs_t}
\end{figure*}

\begin{figure*}[ht!]
    \centering
    \subfigure[Evolution of $\dless{\var{T}}$ for case~3.]
    {\includegraphics[scale=0.35]{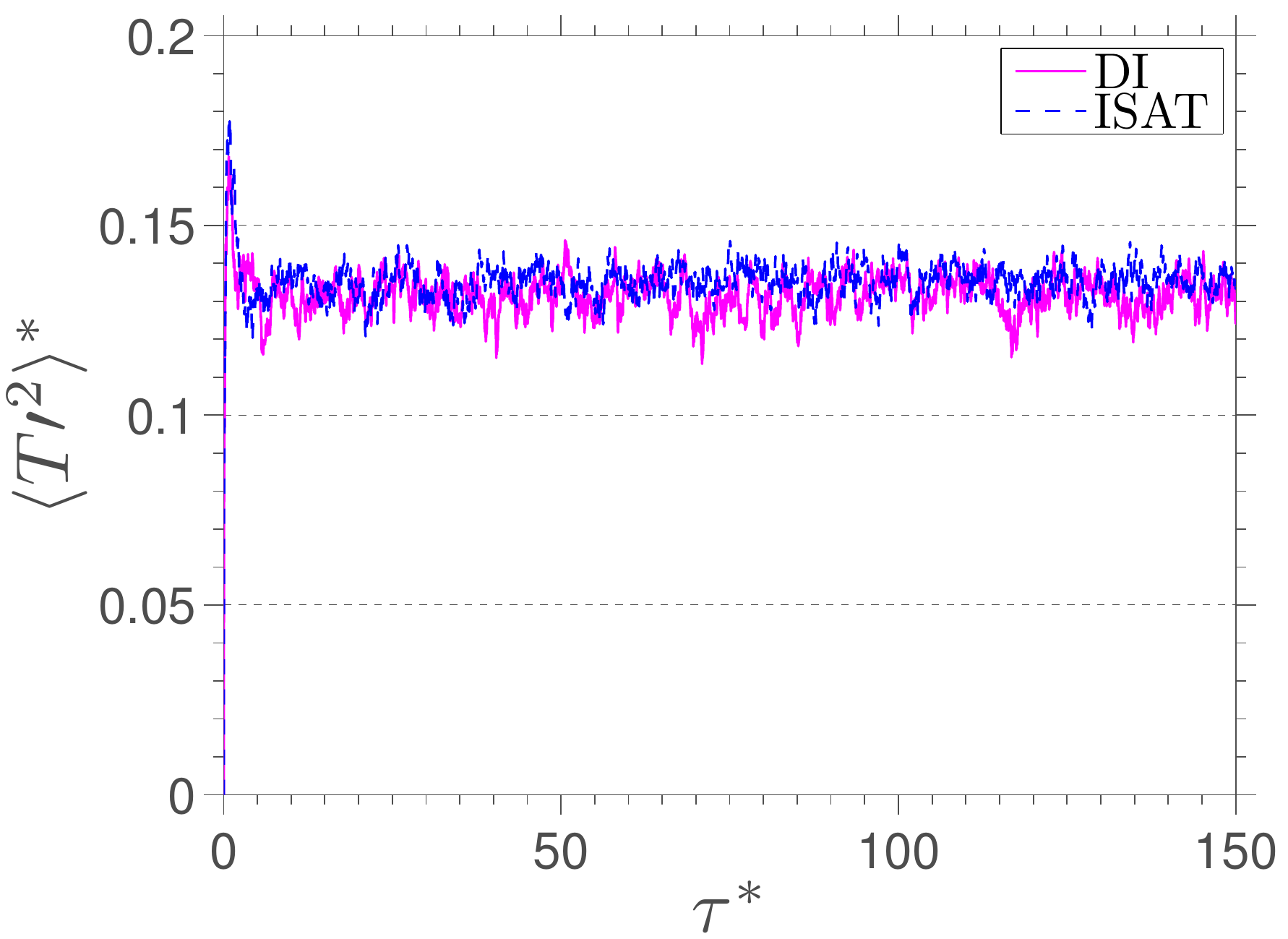}}
    \subfigure[Evolution of $\var{Y_{OH}}$ for case~3.]
    {\includegraphics[scale=0.35]{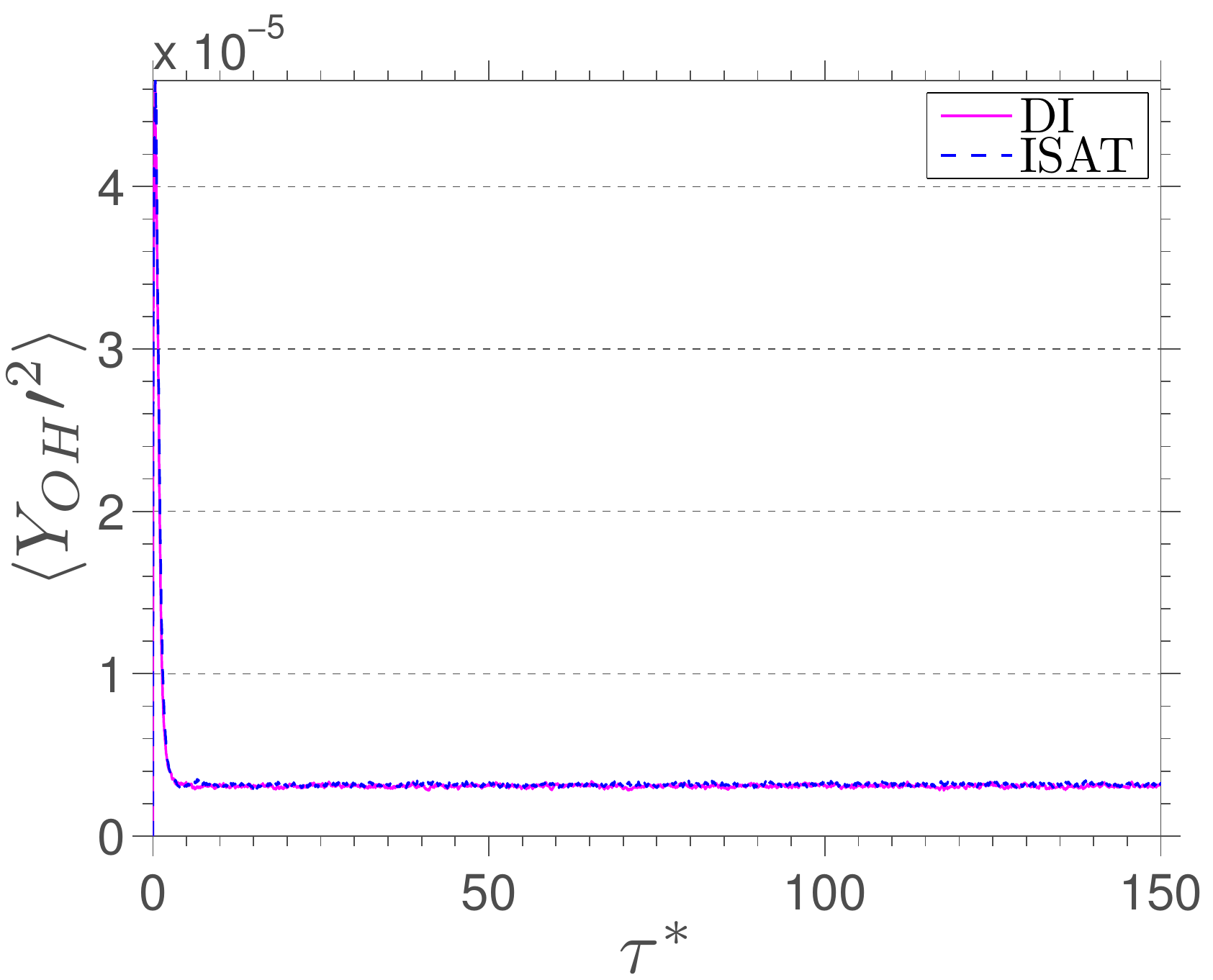}}
    \caption{Comparison between DI (---) and ISAT (- - -) calculations 
    of the ensemble variance of the reduced temperature and $OH$ mass fraction.}
    \label{case3_T_OH_var_vs_t}
\end{figure*}

The comparison between DI and ISAT results for the first two 
statistical moments of $\dless{T}$ and $Y_{OH}$, for case~3,
are presented in Figures~\ref{case3_T_OH_mean_vs_t}~and~\ref{case3_T_OH_var_vs_t}.
One can note reasonable and good agreements for the $\dless{T}$
and the $Y_{OH}$ statistics, respectively. The results of case~3
show a large discrepancy for the $\dless{T}$ than that obtained in case~1. 
The reaction mechanism of methane is much more complex than the 
mechanism used to model the carbon monoxide chemical kinetics in case~1, 
which would in principle lead to the PMSR with methane to assume a wider range 
of possible thermodynamic states. Thus, it could be expected that the present 
binary search tree, with $60\unit{k}$ entries (almost similar to that used in case~1),
would yield comparatively lower accuracy.

\begin{figure*}[ht!]
    \centering
    \subfigure[Histogram of $\dless{T}$ for case~3.]
    {\includegraphics[scale=0.35]{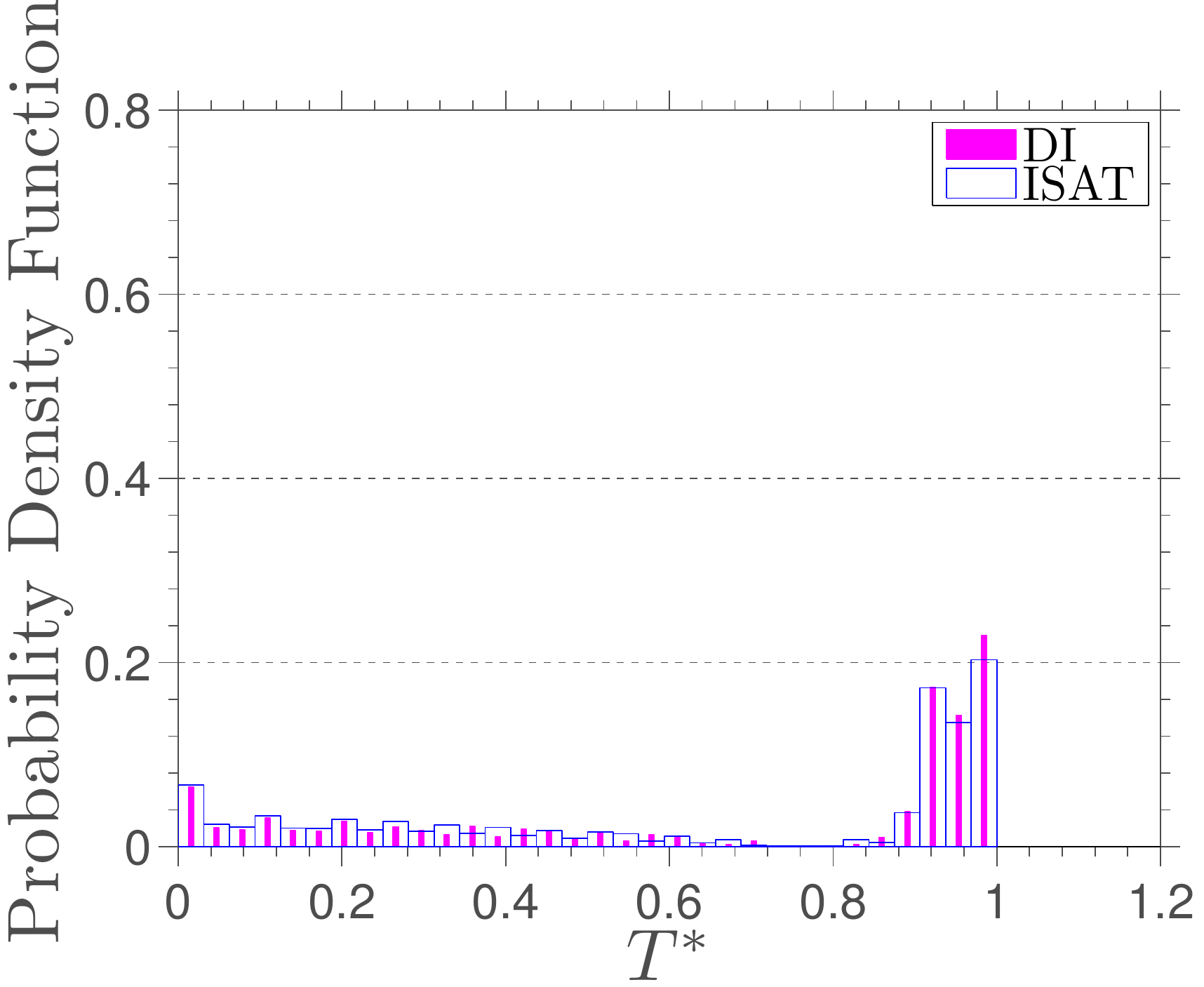}}
    \subfigure[Histogram of $Y_{HCO}$ for case~3.]
    {\includegraphics[scale=0.35]{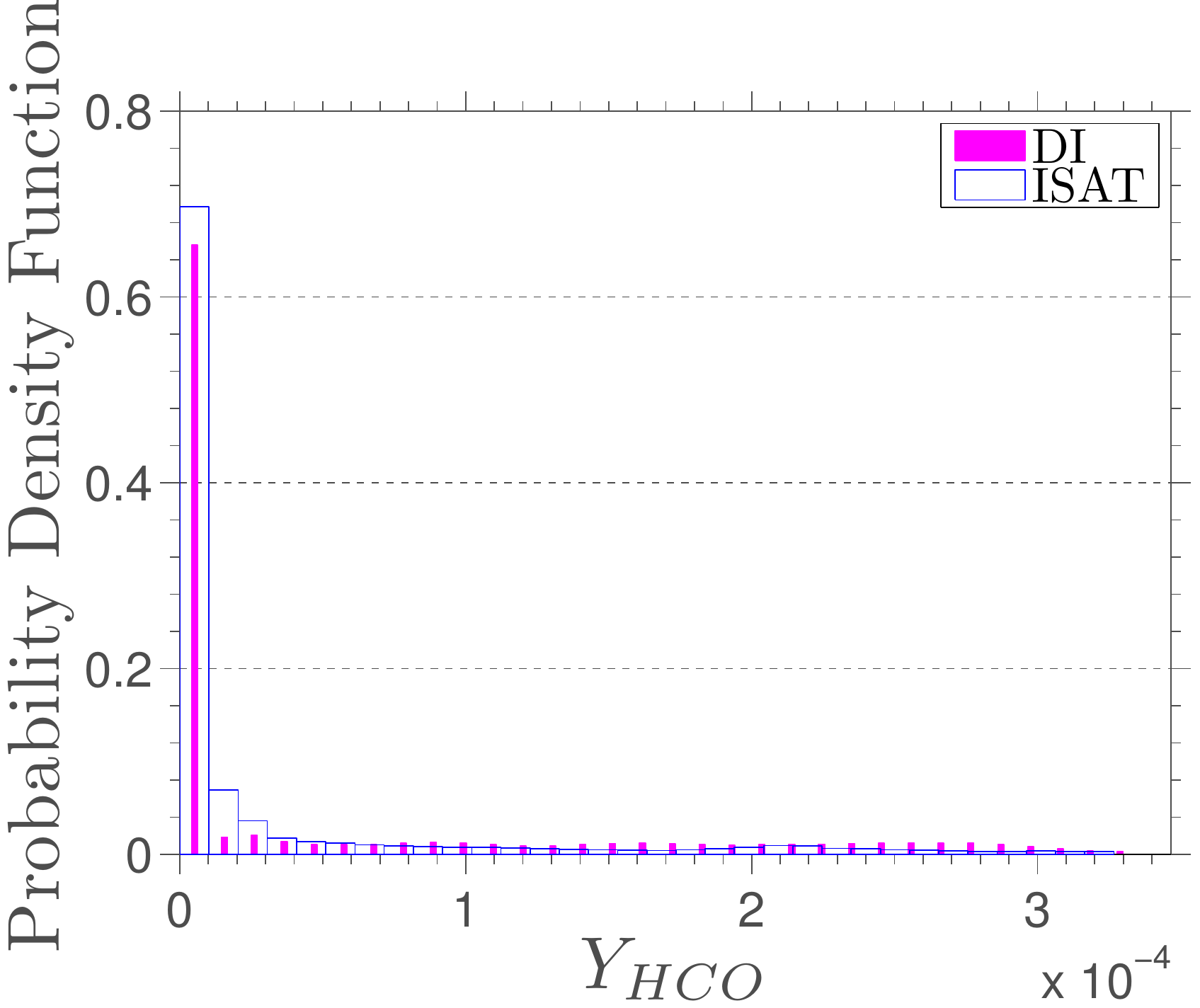}}
    \caption{Comparison between DI and ISAT computations of the 
    mean histograms (averaged over the last $50$ residence times)
    of the reduced temperature and $HCO$ mass fraction.}
    \label{case3_pdf_T_HCO}
\end{figure*}

Finally, Figure~\ref{case3_pdf_T_HCO} presents
the comparison between DI and ISAT computations 
of the mean histograms, averaged over the last $50$ residence times,
of the $\dless{T}$ and $Y_{HCO}$ for case~3.
The temperature histogram presents a bimodal distribution with large
probability of finding $\dless{T} \geq 0.9$ and a broader temperature
distribution leaning to the fresh gases. On the other hand, $Y_{HCO}$
histogram shows a distribution essentially concentrated in the 
fresh gases region, and nearly homogeneous elsewhere. This behavior
illustrates the fact that $HCO$, an intermediate species, appears
in low concentration in the burned gases.

\subsection{Analysis of ISAT accuracy}
\label{analysis_pmsr_accur}

The \emph{relative local error} of $\psi$, 
which is a time dependent function, is defined as

\begin{equation}
		\rerror{\psi}(t) =
		 \Big | \frac{\psi_{DI} (t) - \psi_{ISAT} (t)}{\psi_{DI} (t)} \Big |,
		 \label{def_rlerror}
\end{equation}

\noindent
where subscripts $_{DI}$ and $_{ISAT}$ denote DI and ISAT 
calculations of $\psi$, respectively. This metric represents a local measure 
of error incurred by ISAT.

\begin{figure*}[ht!]
    \centering
    \subfigure[Evolution of $\rerror{ \dless{\mean{T}} }$ for case~1]
    {\includegraphics[scale=0.35]{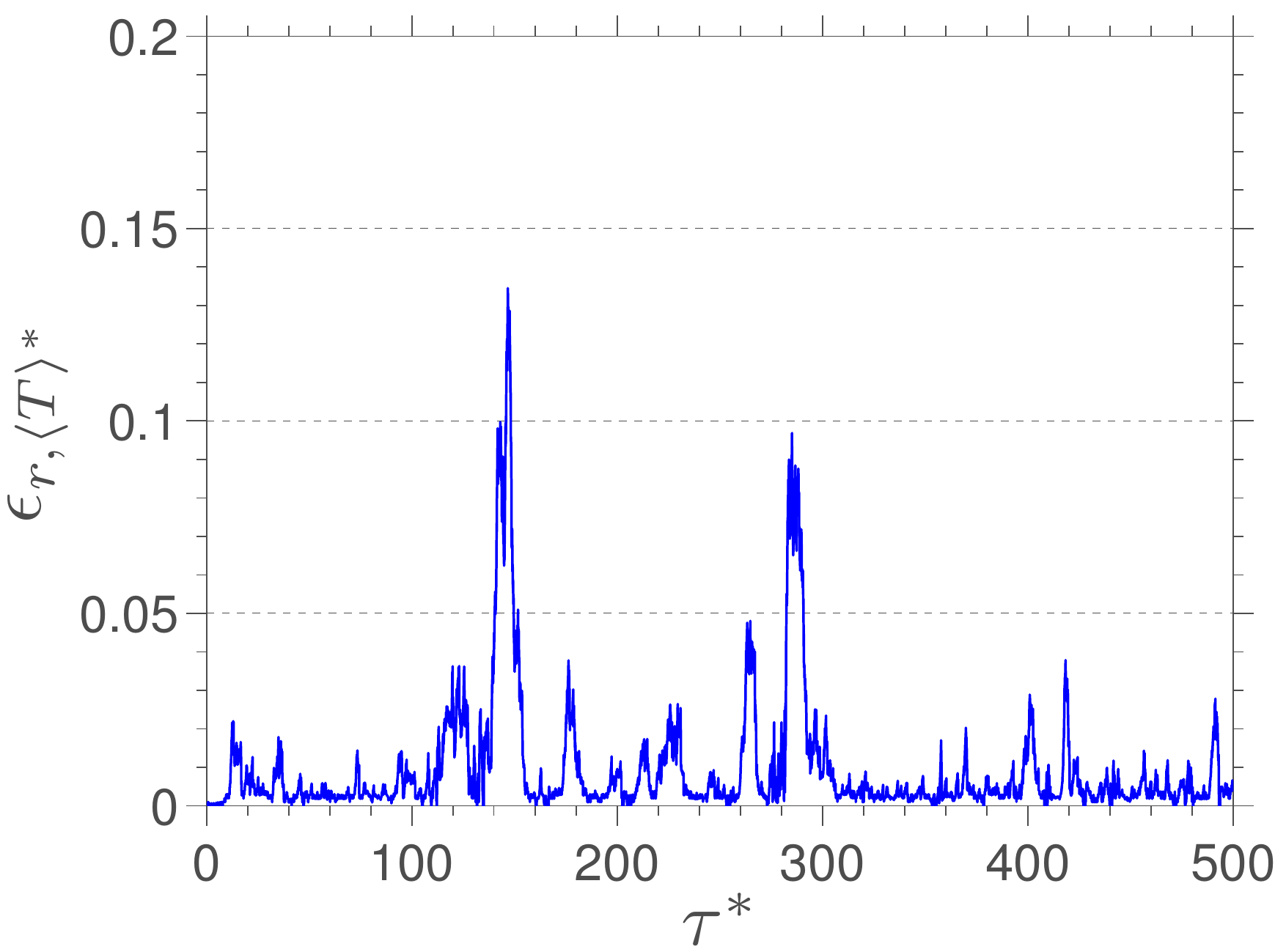}}
    \subfigure[Evolution of $\rerror{ \dless{\mean{T}} }$ for case~2]
    {\includegraphics[scale=0.35]{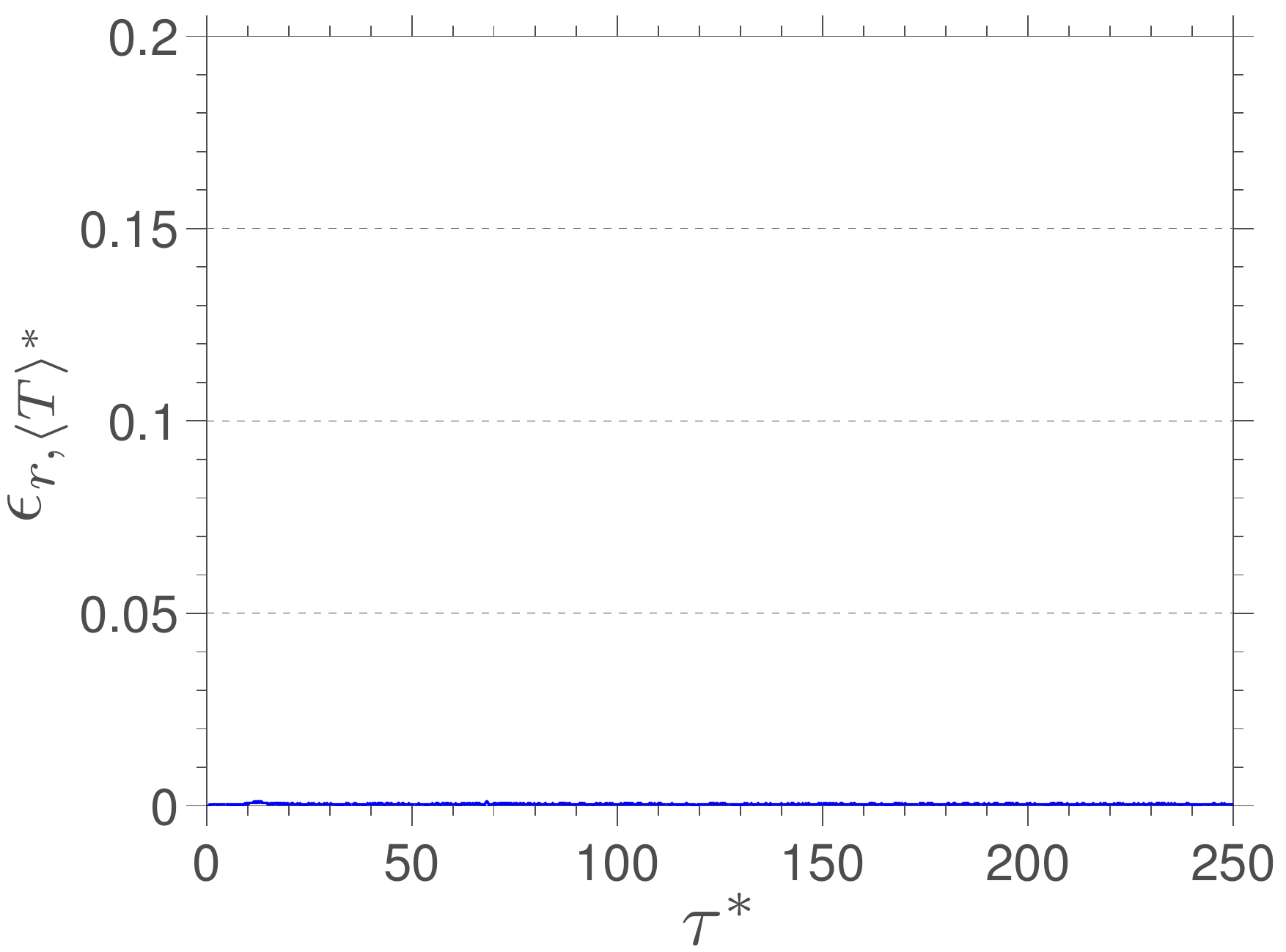}}\\
    \subfigure[Evolution of $\rerror{ \mean{Y_{O}} }$ for case~1]
    {\includegraphics[scale=0.35]{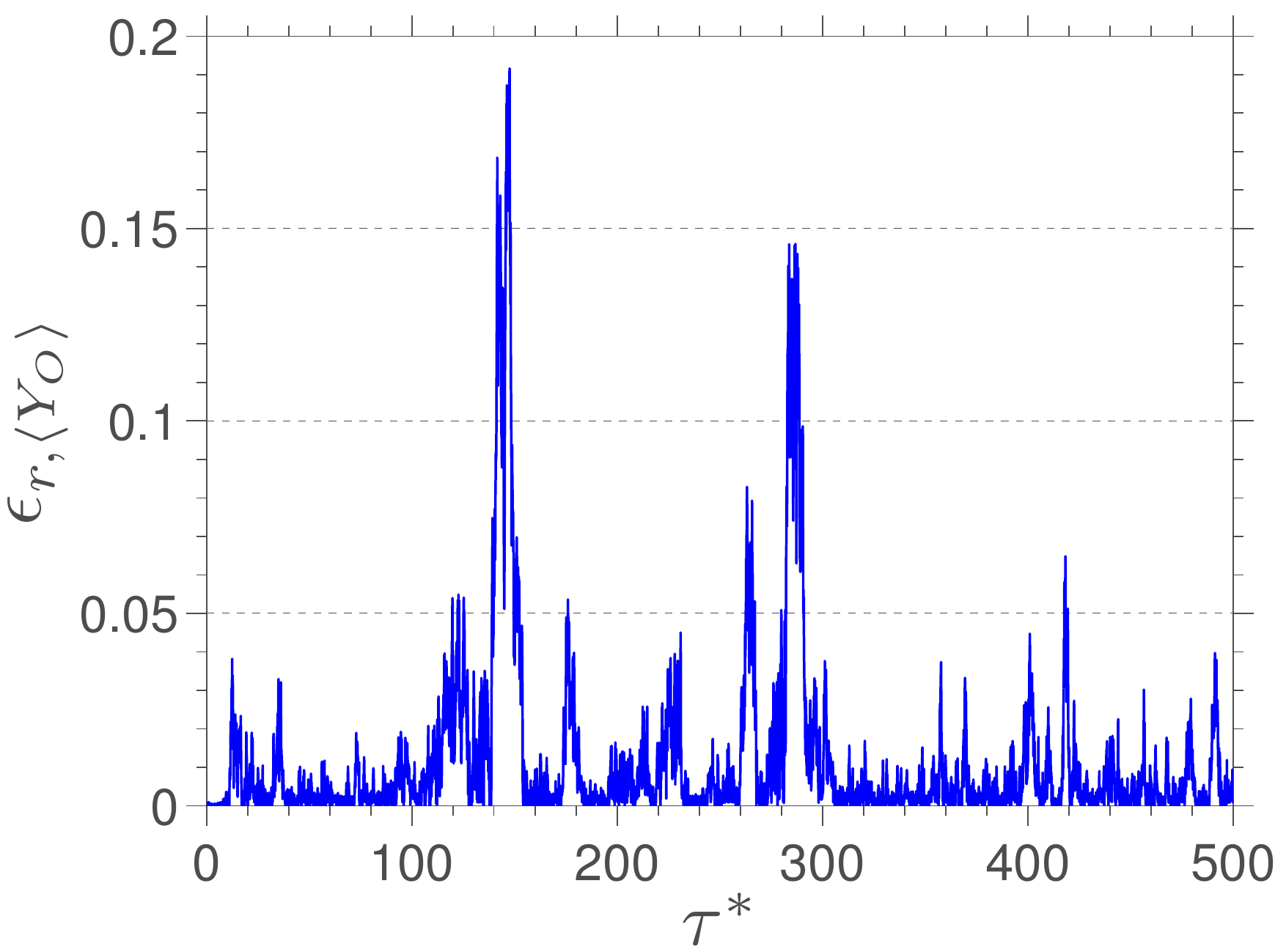}}
    \subfigure[Evolution of $\rerror{ \mean{Y_{O}} }$ for case~2]
    {\includegraphics[scale=0.35]{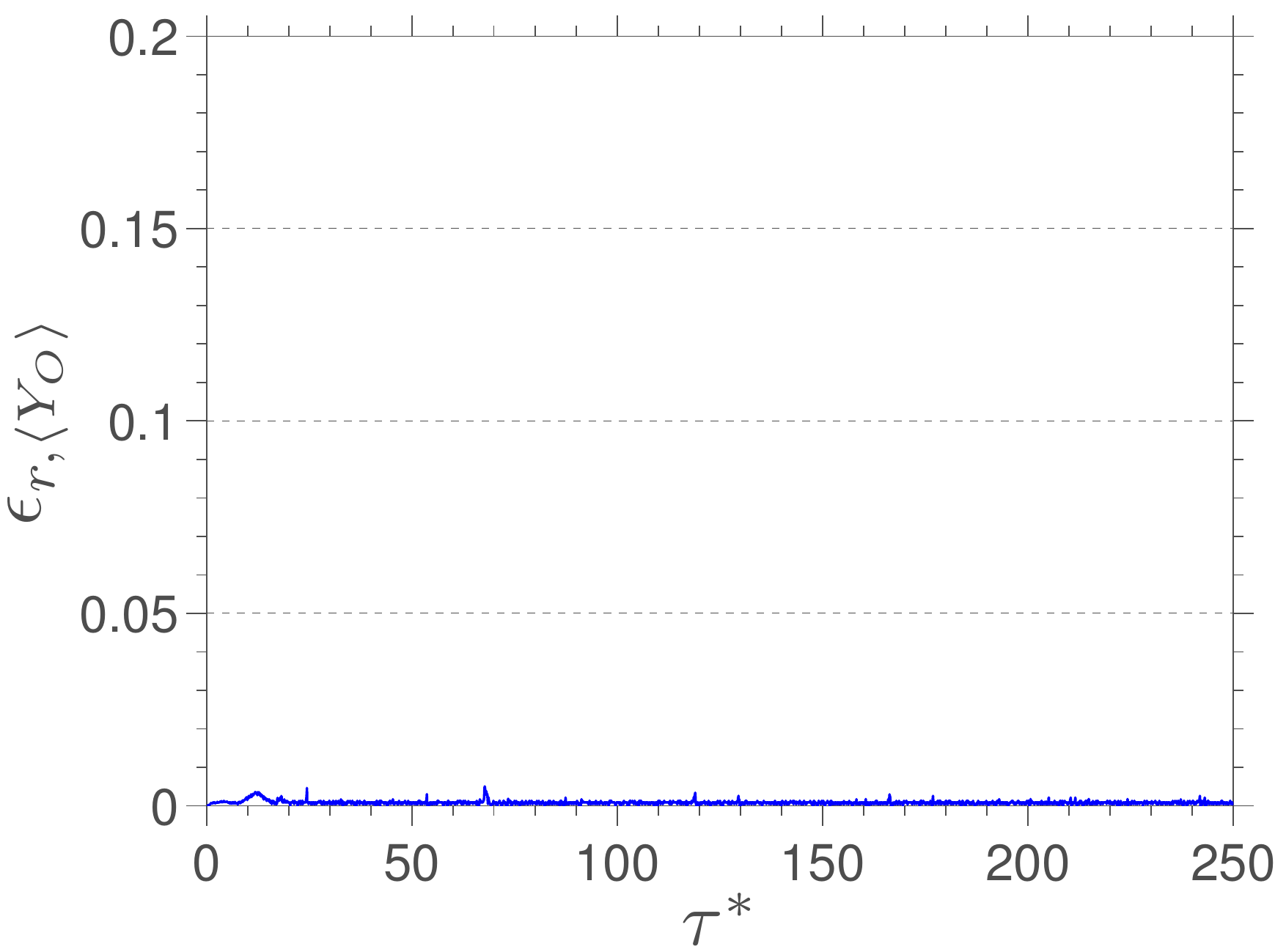}}\\
    \caption{Evolution of the relative local error for the ensemble average of the
    reduced temperature and of the $O$ mass fraction, using the
    same statistical seeds in both cases.}
    \label{cases12_error_T_O_mean_vs_t}
\end{figure*}

\begin{figure*}[ht!]
    \centering
    \subfigure[Evolution of $\rerror{ \dless{\var{T}} }$ for case~1]
    {\includegraphics[scale=0.35]{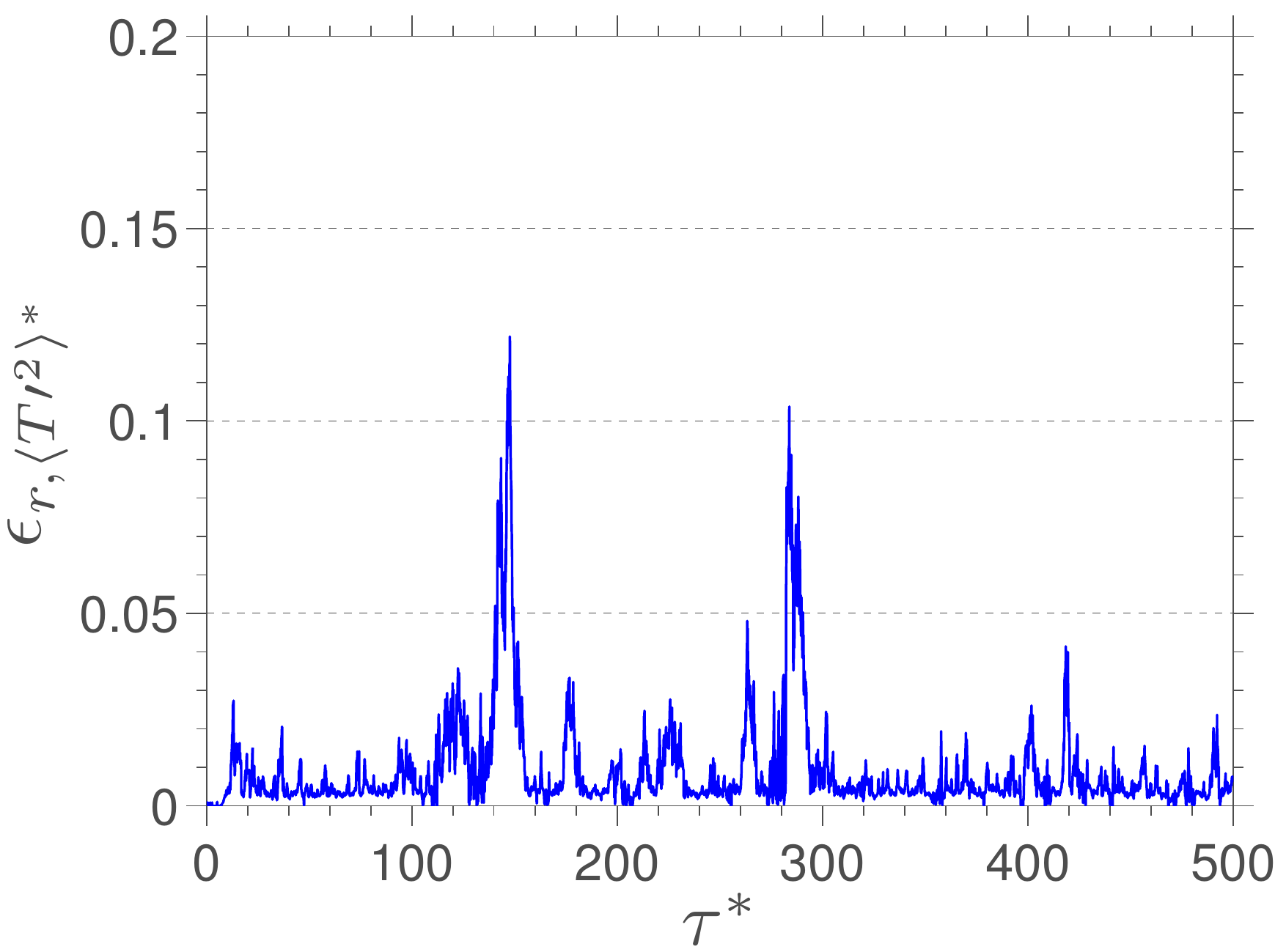}}
    \subfigure[Evolution of $\rerror{ \dless{\var{T}} }$ for case~2]
    {\includegraphics[scale=0.35]{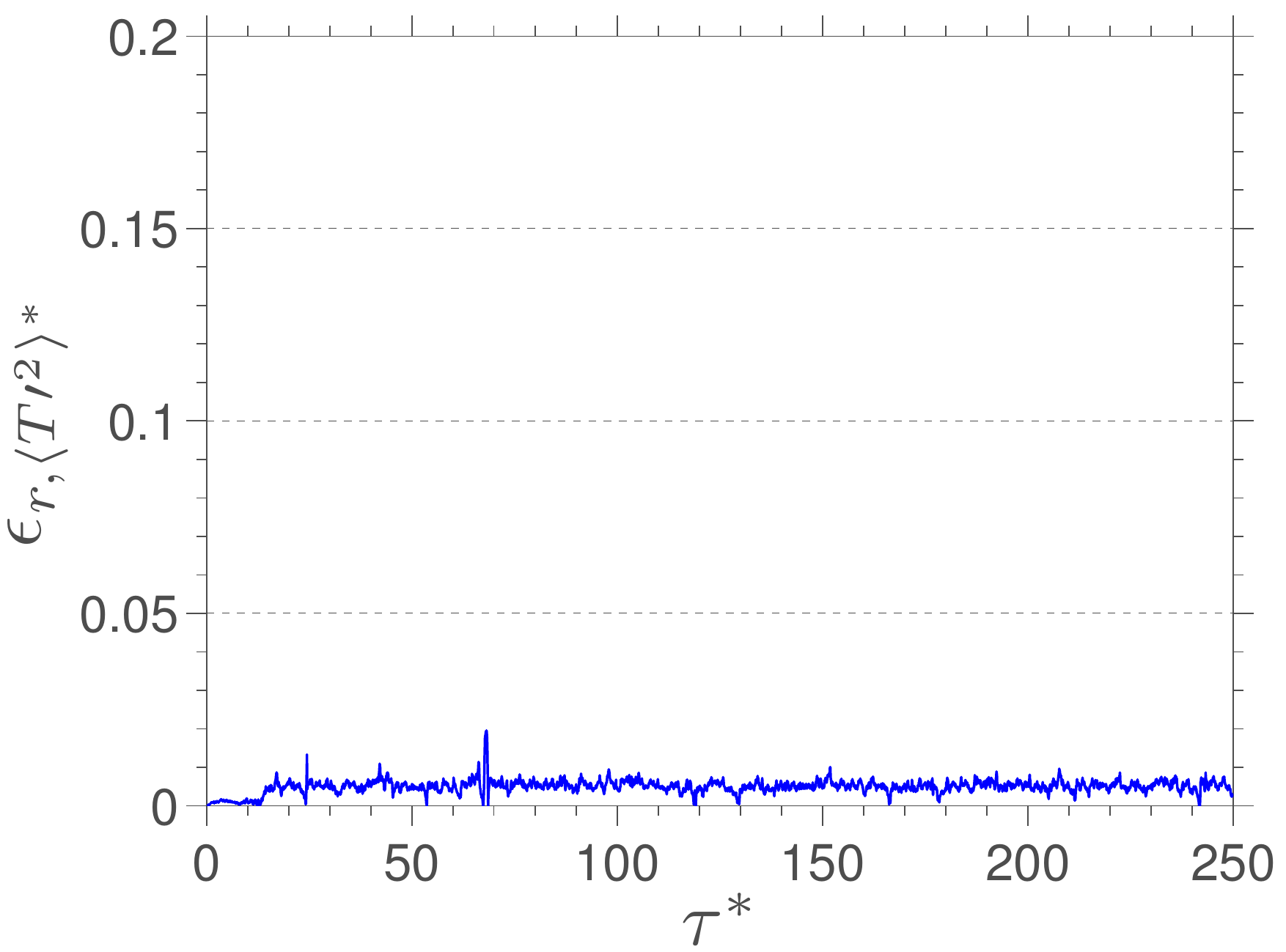}}\\
    \subfigure[Evolution of $\rerror{ \var{Y_{O}} }$ for case~1]
    {\includegraphics[scale=0.35]{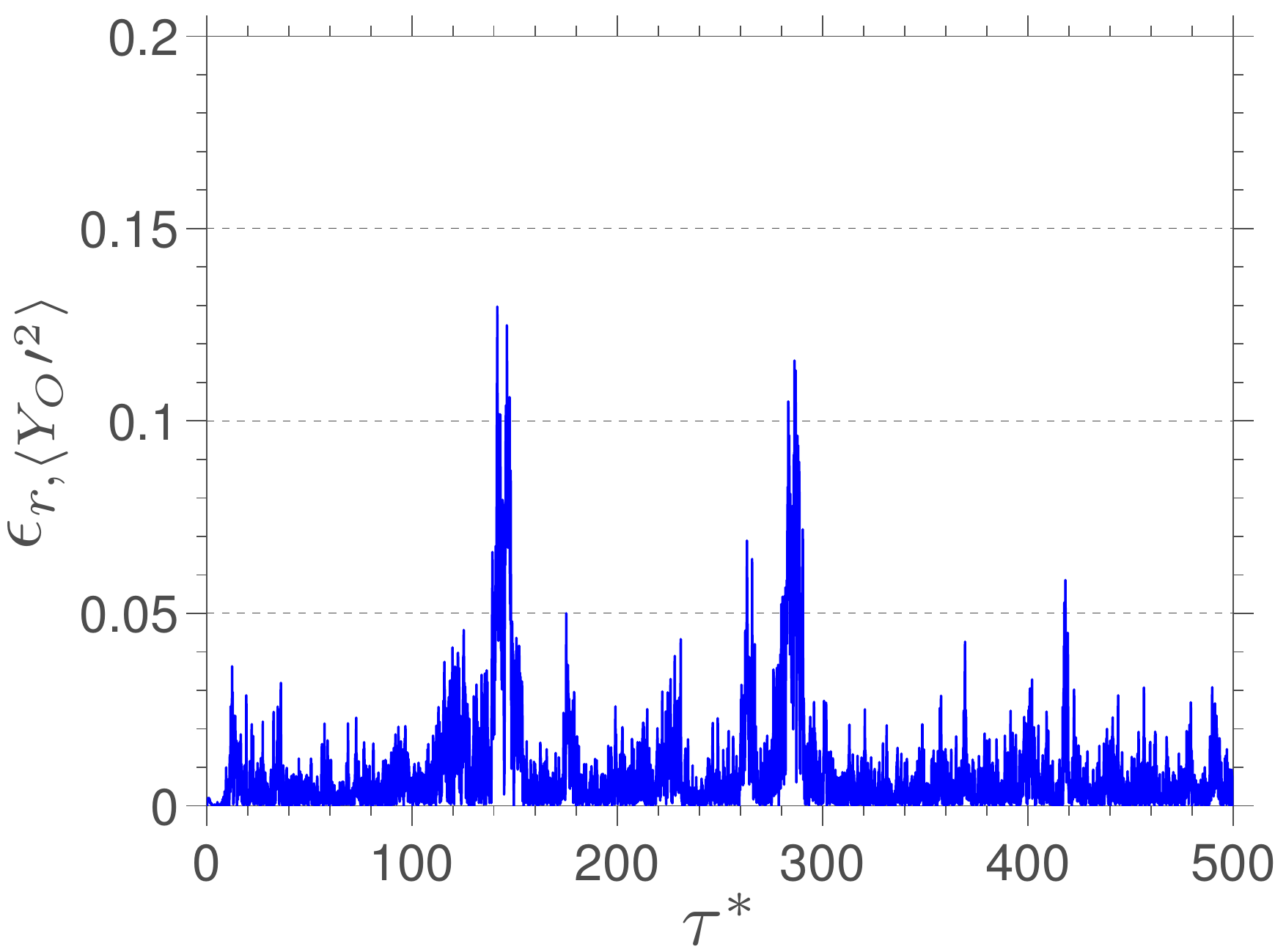}}
    \subfigure[Evolution of $\rerror{ \var{Y_{O}} }$ for case~2]
    {\includegraphics[scale=0.35]{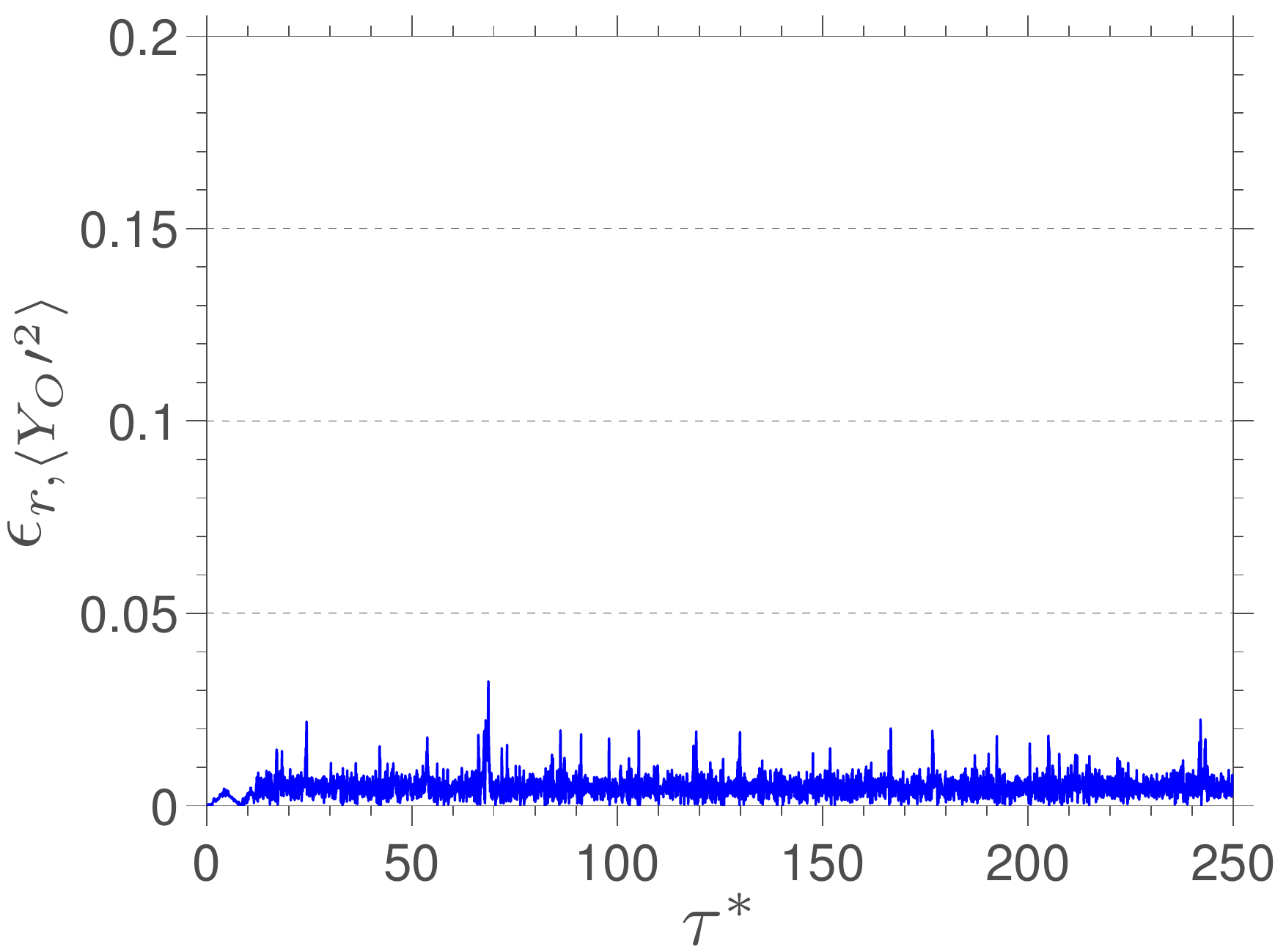}}\\
    \caption{Evolution of the relative local error for the ensemble variance of the
    reduced temperature and of the $O$ mass fraction, using the
    same statistical seeds in both cases.}
    \label{cases12_error_T_O_var_vs_t}
\end{figure*}

The evolution of the relative local errors of $\dless{T}$ and $Y_{O}$ 
statistical moments are presented in
Figures~\ref{cases12_error_T_O_mean_vs_t}~and~\ref{cases12_error_T_O_var_vs_t}.
Concerning case~1 errors, one can observe a large 
statistical variation due to the stochastic nature of the PMSR model, 
with amplitudes reaching 13\%. In case~2, relative errors of 
the order of 1\% only can be observed. 

\begin{figure*}[ht!]
    \centering
    \subfigure[Evolution of $\dless{\mean{T}}$ for case~1.]
    {\includegraphics[scale=0.35]{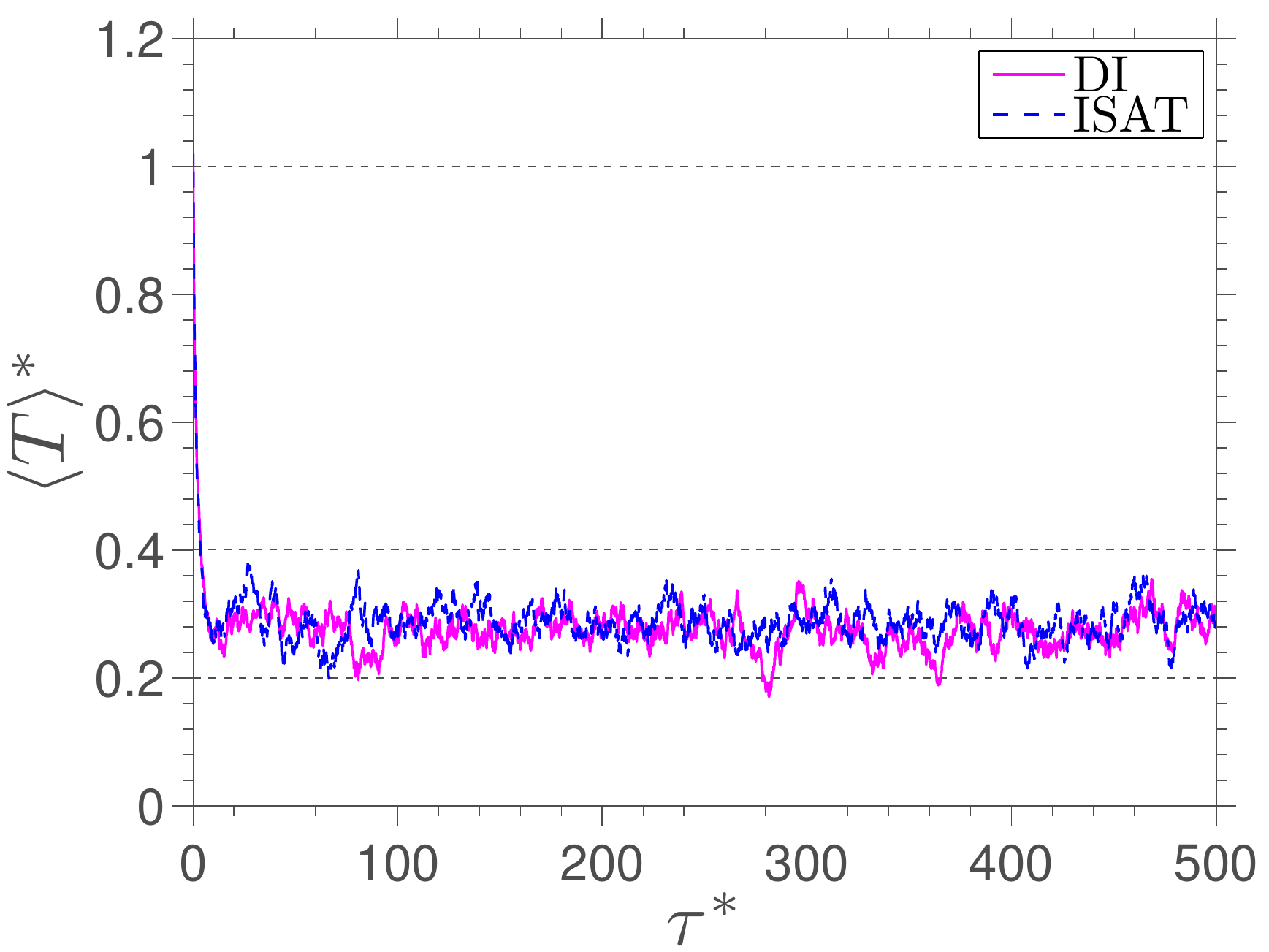}}
    \subfigure[Evolution of $\dless{\mean{T}}$ for case~2.]
    {\includegraphics[scale=0.35]{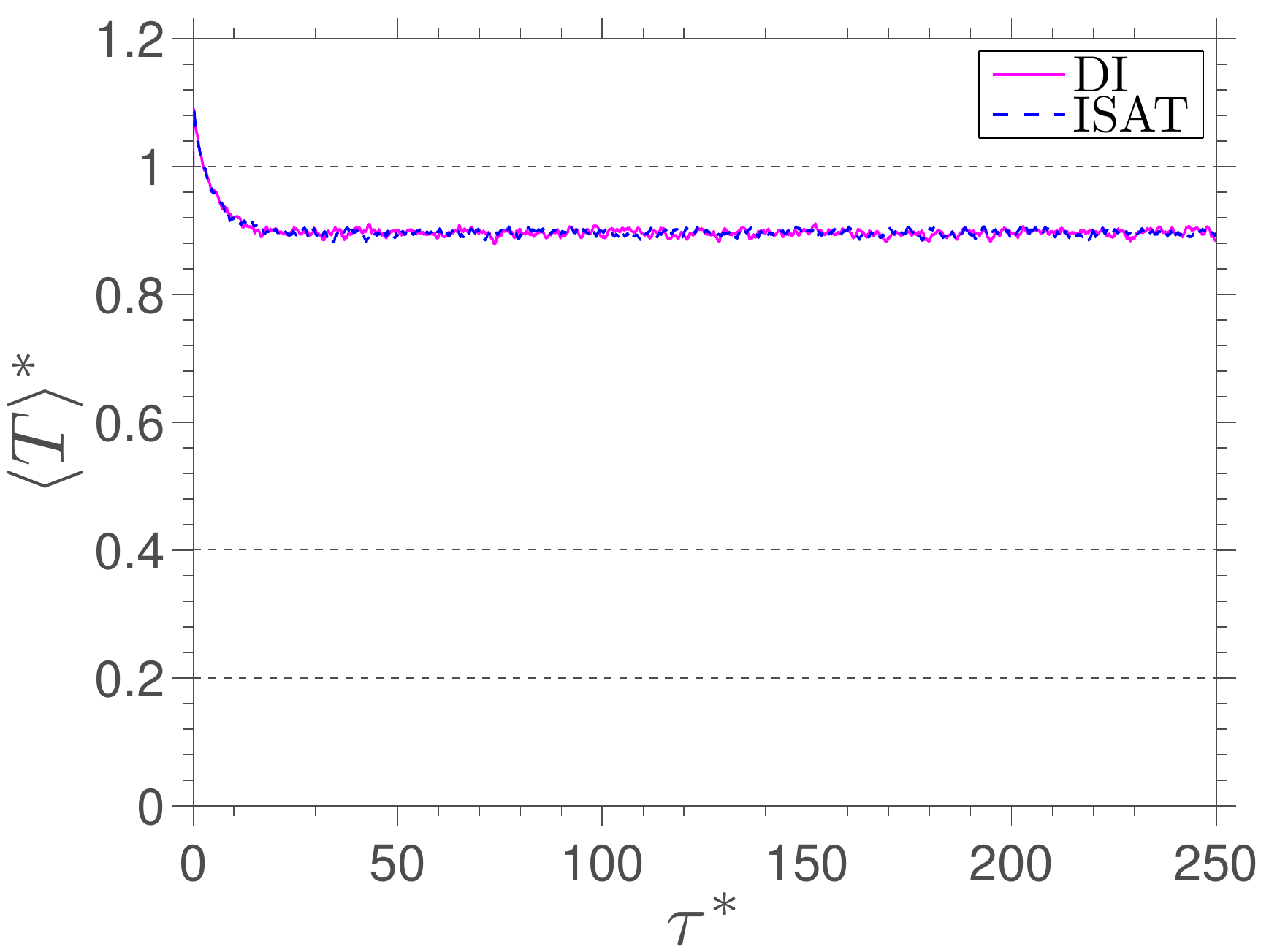}}\\
    \subfigure[Evolution of $\rerror{ \dless{\mean{T}} }$ for case~1]
    {\includegraphics[scale=0.35]{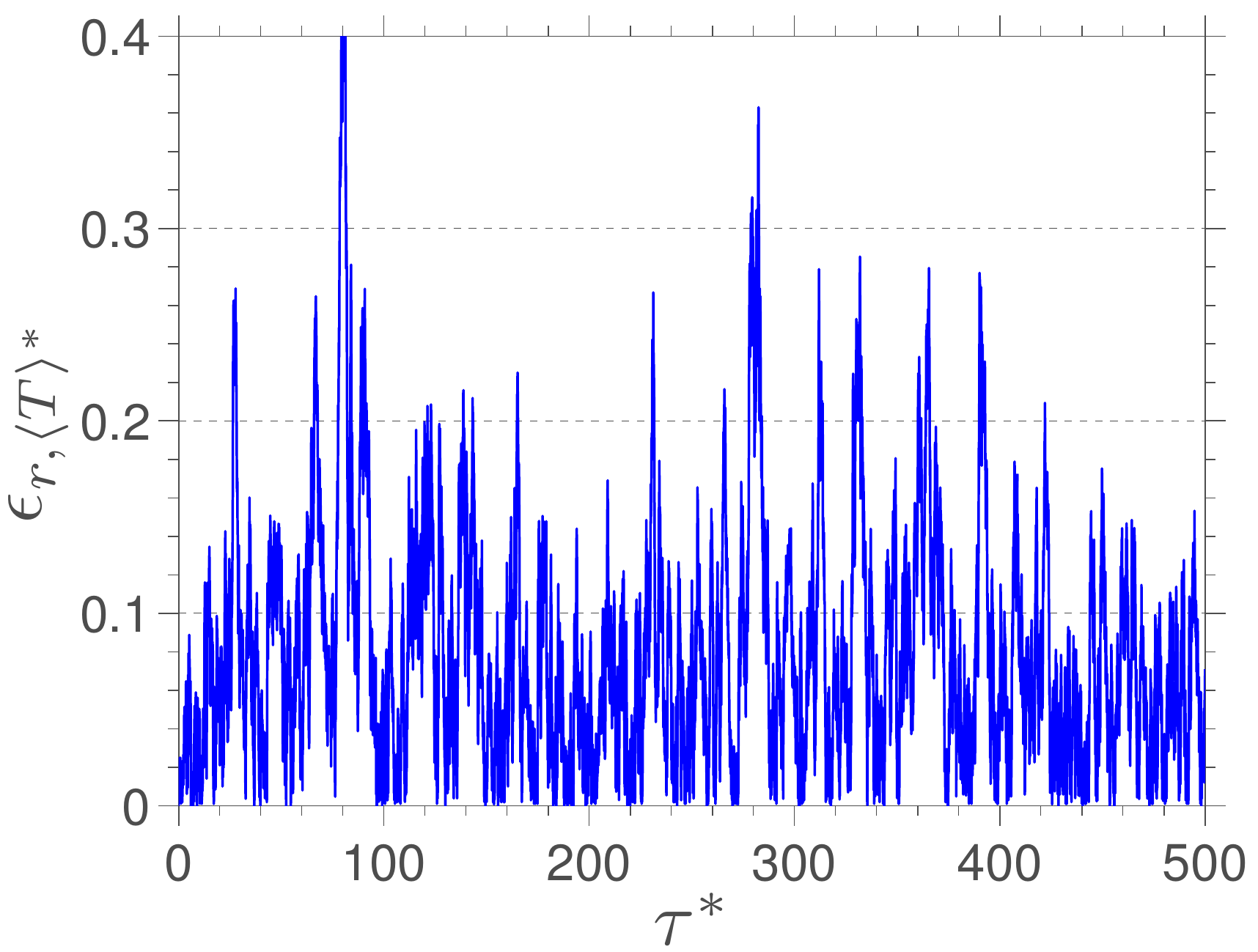}}
    \subfigure[Evolution of $\rerror{ \dless{\mean{T}} }$ for case~2]
    {\includegraphics[scale=0.35]{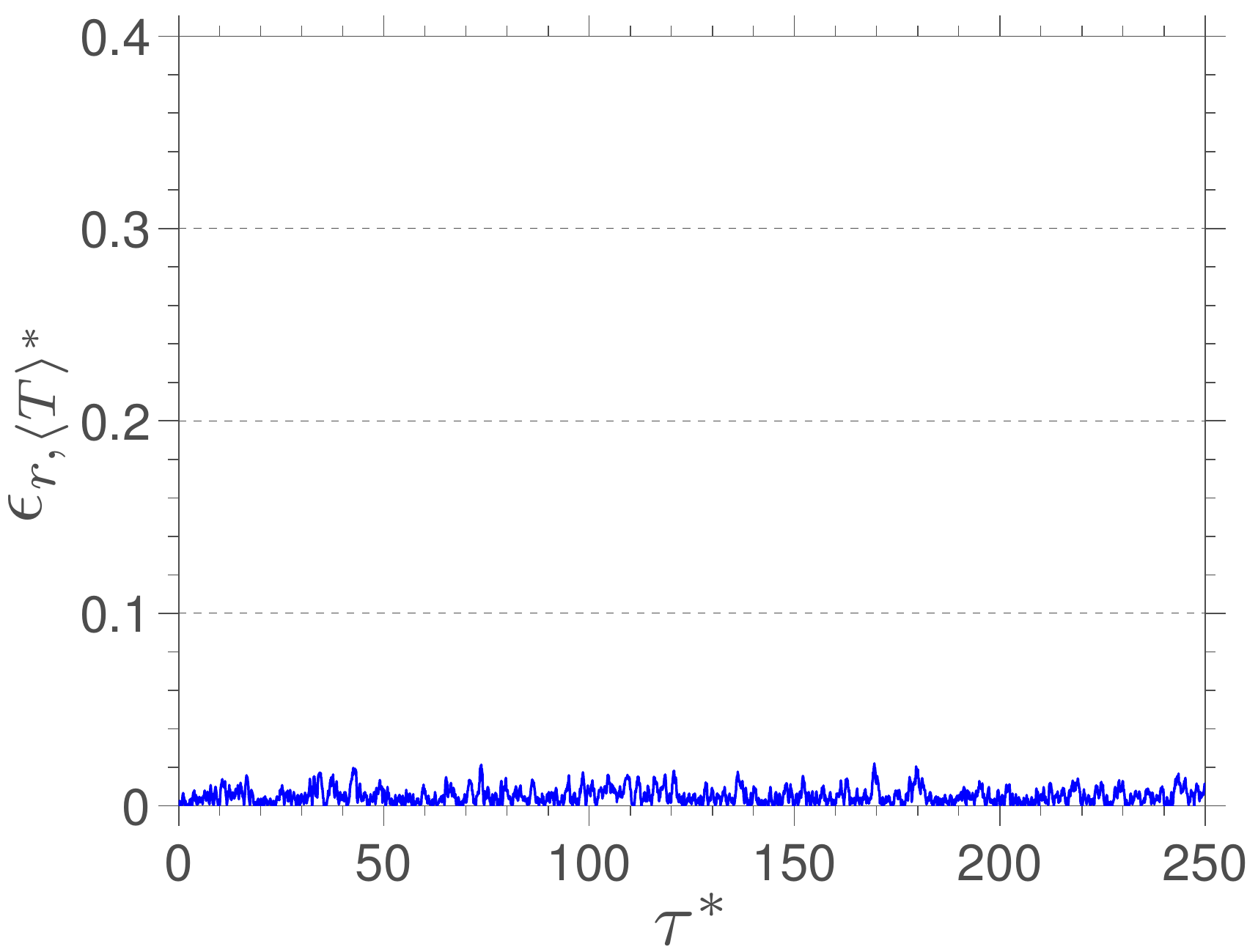}}\\
    \caption{Comparison between DI (---) and ISAT (- - -) calculations, 
    using different statistical seeds, of the ensemble average of the reduced temperature 
    and the corresponding relative local errors.}
    \label{cases12_T_mean_error_vs_t_ds}
\end{figure*}

All realizations of the stochastic processes in DI and ISAT 
calculations shown above use the same \emph{statistical seed} 
for pseudorandom number generator.
However, since the mixing model is statistical in nature, 
it could be expected that the seed value may influence the 
ISAT behavior. Therefore, if the DI calculation seed is kept 
fixed and ISAT seed is changed, the results for 
$\dless{\mean{T}}$ and $\rerror{ \dless{\mean{T}} }$ are 
expected to be modified.  This can be seen in 
Figure~\ref{cases12_T_mean_error_vs_t_ds}, 
where one may observe, by comparison with 
Figures~\ref{cases12_T_mean_vs_t}~and~\ref{cases12_error_T_O_mean_vs_t},
an increase in the $\rerror{ \dless{\mean{T}} }$ for both cases.

\begin{figure*}[ht!]
    \centering
    \subfigure[Histogram of $\dless{T}$ for case~1.]
    {\includegraphics[scale=0.35]{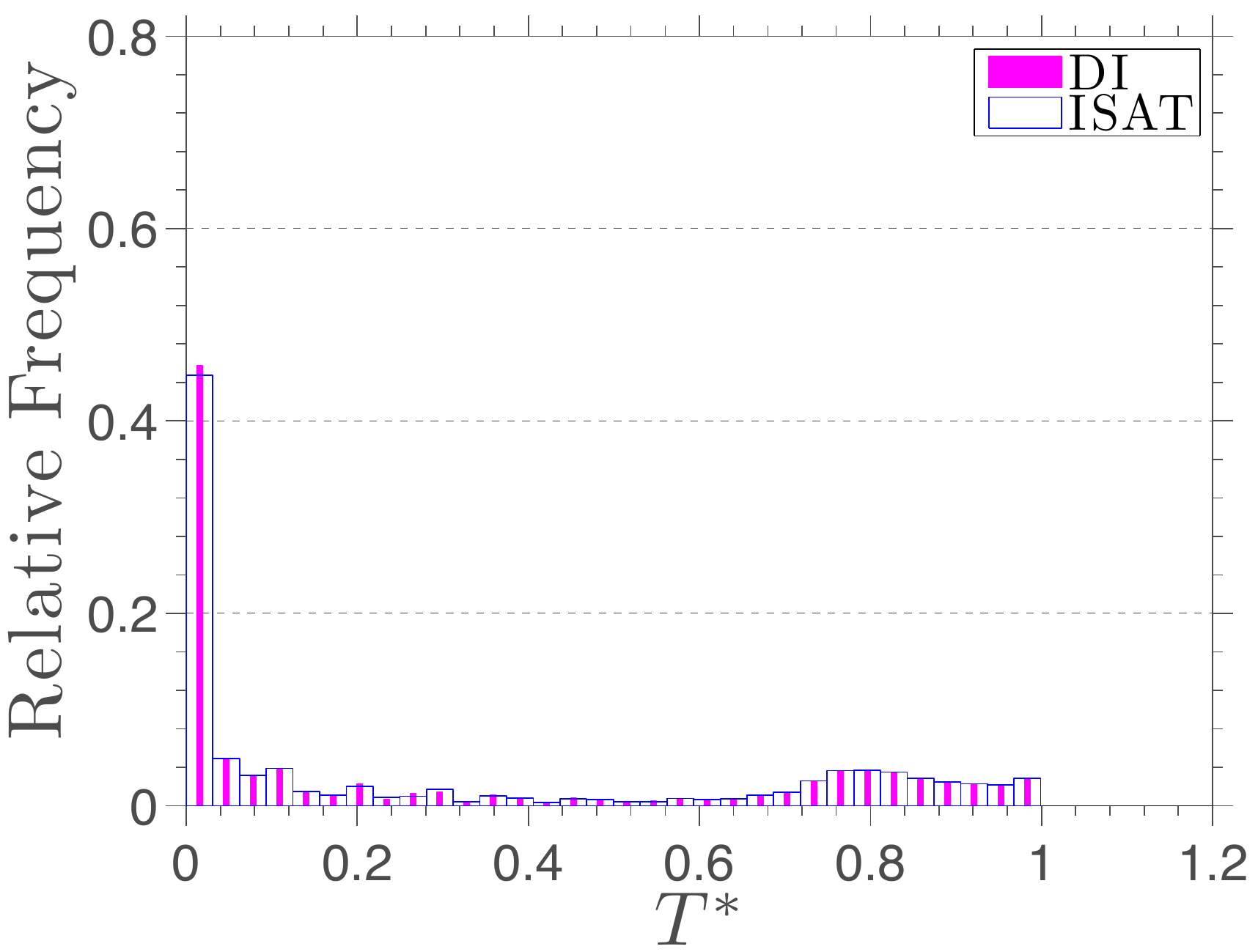}}
    \subfigure[Histogram of $\dless{T}$ for case~2.]
    {\includegraphics[scale=0.35]{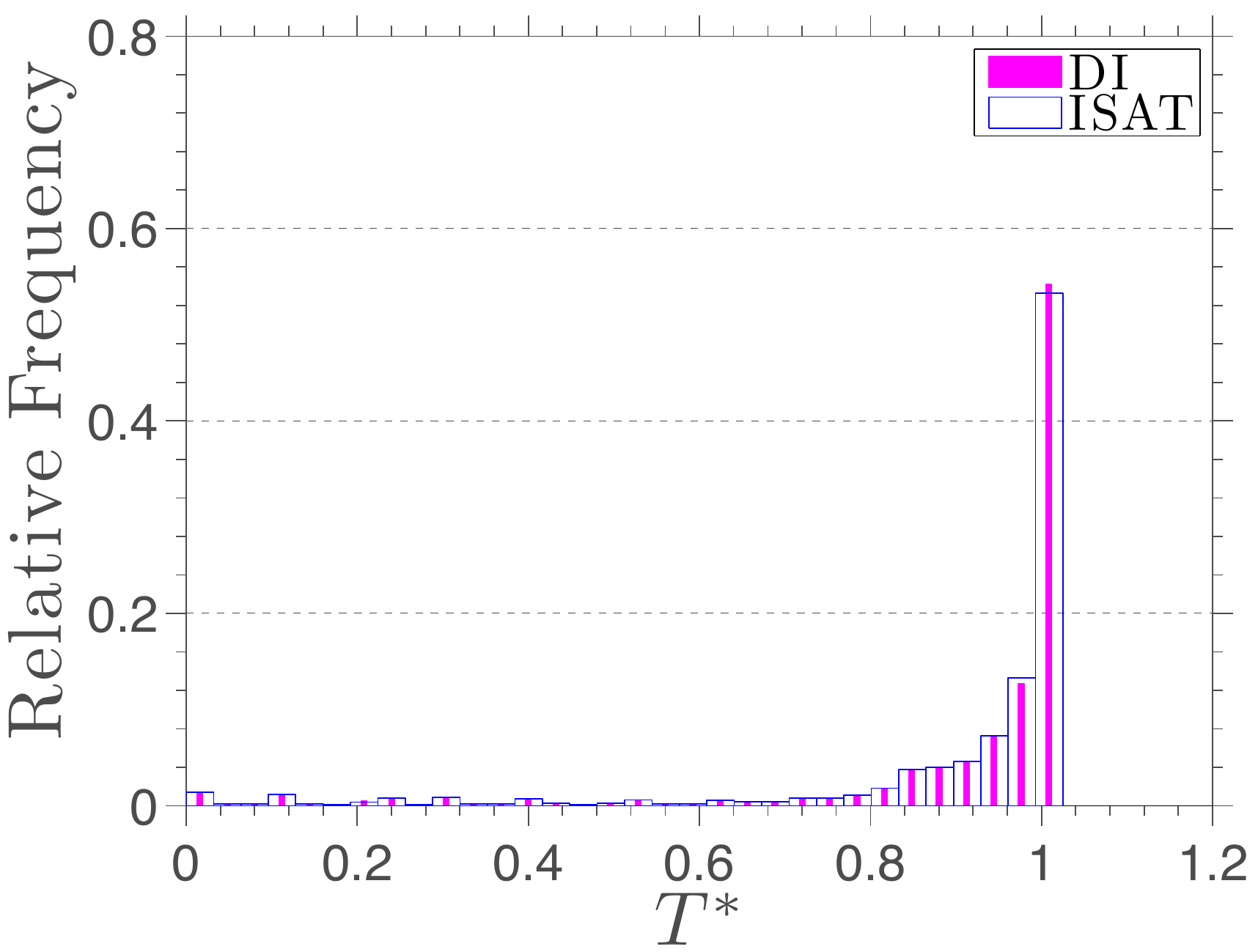}}
    \caption{Comparison between DI and ISAT calculations,
    using different statistical seeds, of the mean histograms 
    (averaged over the last $50$ residence times)
    of the reduced temperature.}
    \label{cases12_pdf_T_ds}
\end{figure*}

The analysis of the graphs in Figure~\ref{cases12_T_mean_error_vs_t_ds}
indicates that 1024 particles are not sufficient to guarantee the statistical 
independence of the computed results. This hypothesis is also confirmed if
one considers Figure~\ref{cases12_pdf_T_ds}, where it is possible
to see discrepancies in DI and ISAT mean histograms of $\dless{T}$ 
for both cases, even if, qualitatively these histograms  are not much 
different than those shown in Figure~\ref{cases12_pdf_T}.
Thus, one can conclude that although the results not exhibit
the statistical independence with a sample of 1024 particles, they
do not vary much with a sample of this size. Possibly a sample of 
4096 particles is sufficient to ensure the independence of the results.

The \define{mean relative error} of $\psi$ over an interval 
$\Delta \tau$, defined as

\begin{equation}
		\funcM{ \rerror{\psi} } =
		\frac{1}{\Delta \tau}
		\int_{t}^{t+\Delta \tau} \rerror{\psi}(t') d t',
		 \label{def_meanrerror}
\end{equation}

\noindent
is a metric that gives a global measure of the error incurred by ISAT.

The mean relative errors associated to the statistical moments 
of the $\dless{T}$ and the $Y_{O}$ are presented in 
Table~\ref{pmsr_errors}.
One can observe that, when the same seed is used for ISAT and DI, 
the mean error of the average properties is smaller than 
1.2\% in case~1, and of 0.1\% in case~2 only. For $\dless{T}$ 
and $Y_{O}$ ensemble variances, the mean error is 
large in case~1. Concerning the simulations where the used
seeds are different, the mean error of the properties average/variance 
are an order of magnitude greater than the previous one.

\begin{table}[ht!]
	\centering
	\caption{Mean relative error of $\psi$ for cases~1~and~2.}
	\input{tabs/pmsr_errors.tab}
	\label{pmsr_errors}
\end{table}

\begin{figure*}[ht!]
    \centering
    \subfigure[$\varepsilon_{g}$ vs $\epstol$ for case~1.]
    {\includegraphics[scale=0.35]{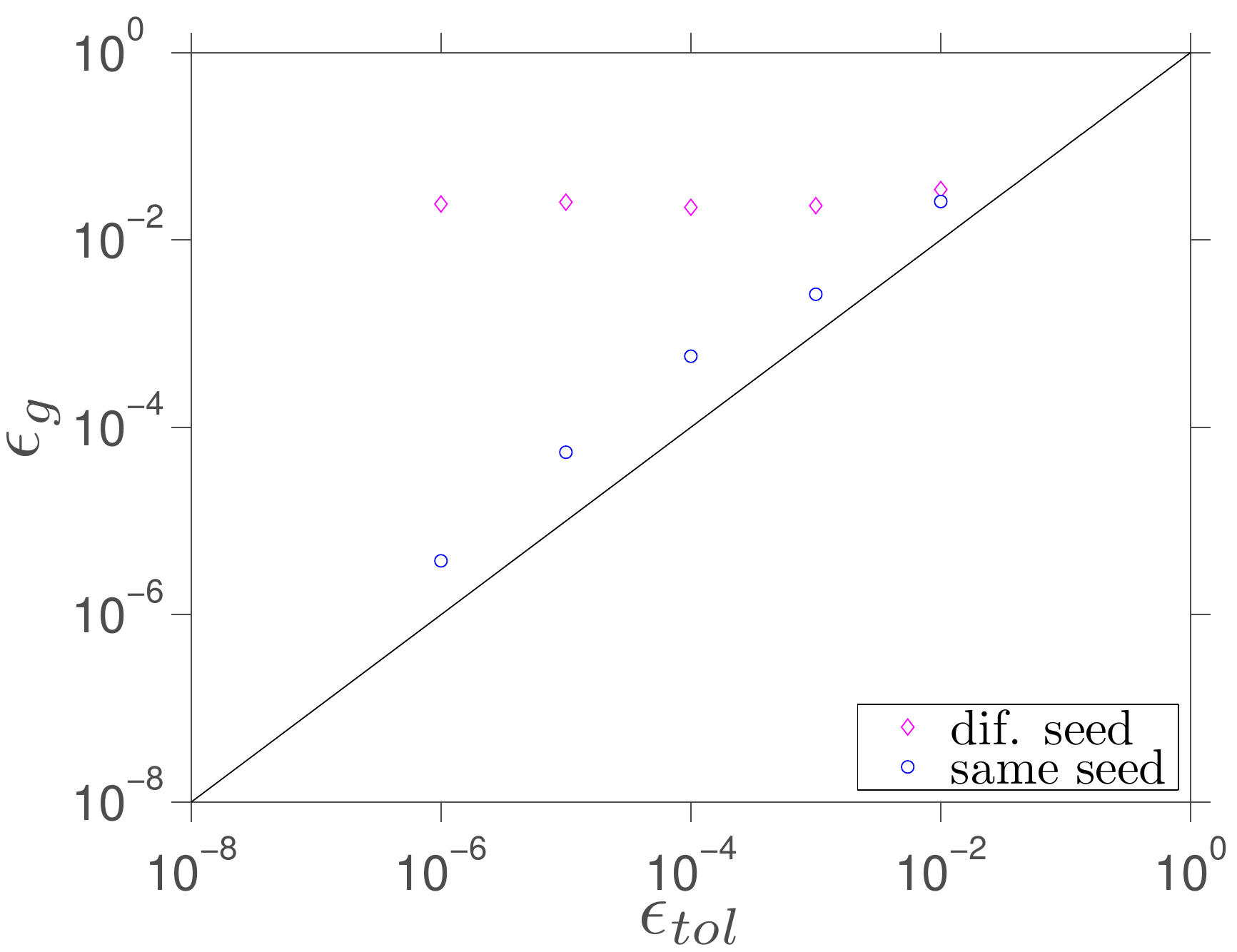}}
    \subfigure[$\varepsilon_{g}$ vs $\epstol$ for case~2.]
    {\includegraphics[scale=0.35]{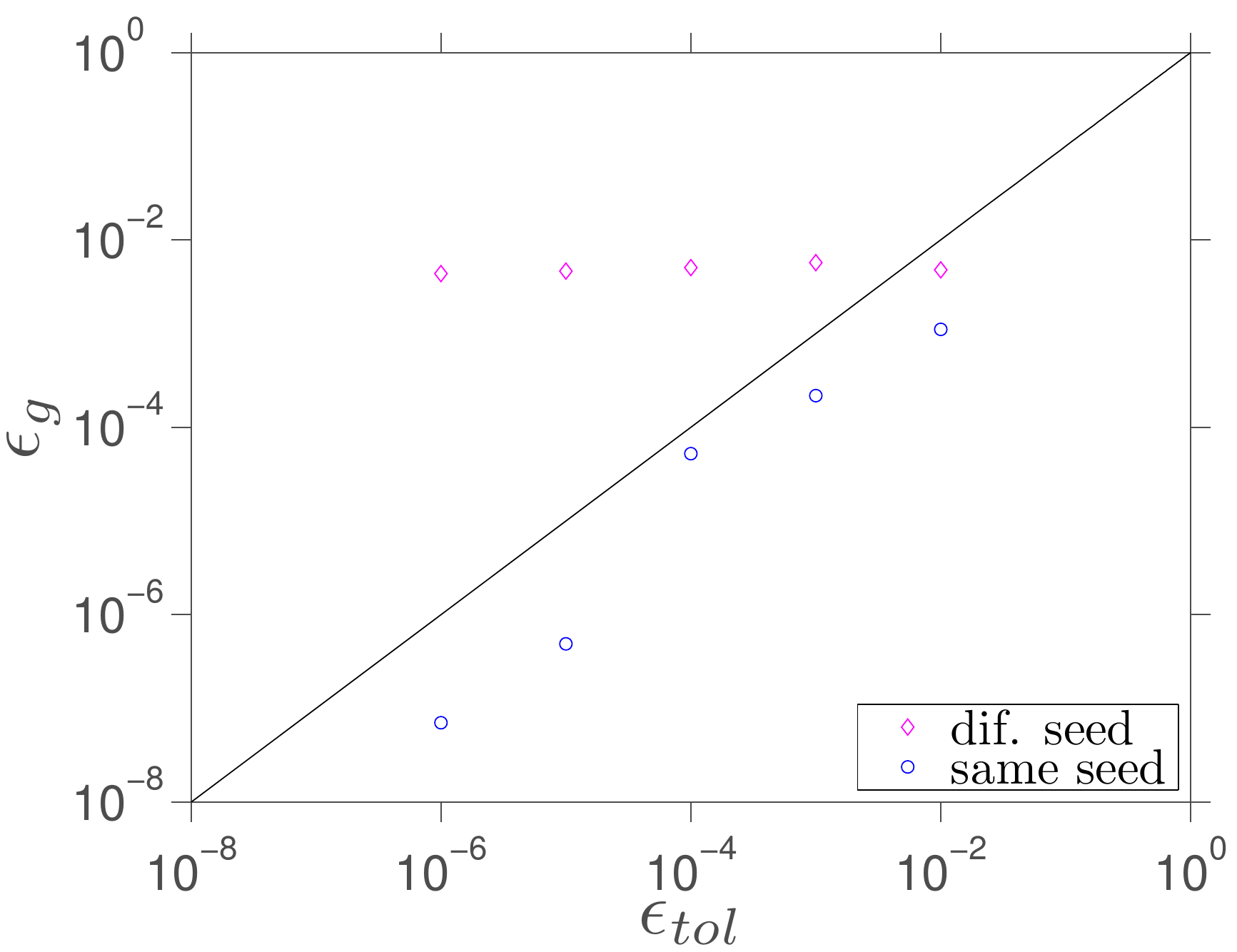}}\\
    \caption{Absolute global errors as function of the error tolerance,
    using a binary search tree with $50\unit{k}$ entries.}
    \label{CO_O2_abs_global_error_50k}
\end{figure*}

In the early development of ISAT technique \cite{pope1997p41}
it was noted that the choice of the tolerance could affect the
accuracy of the problem solution. In order to investigate the
effect of the tolerance on the present results,
Figure~\ref{CO_O2_abs_global_error_50k} presents the
\emph{absolute global error}, $\varepsilon_{g}$,
as a function of ISAT error tolerance for cases~1~and~2.
The absolute global error is defined, over a time interval 
$\Delta \tau$, as

\begin{equation}
		\varepsilon_{g} =
		\frac{1}{\Delta \tau}
		\int_{t}^{t+\Delta \tau}
		\norm{ \mean{\bm{\phi}} (t')_{DI} - \mean{\bm{\phi}} (t')_{ISAT} } ~ dt',
\end{equation}

\noindent
where $\mean{\bm{\phi}}$ denotes a vector which the components
are the ensemble averages of $\bm{\phi}$ components.

As may be seen in Figure~\ref{CO_O2_abs_global_error_50k},
when the same statistical seed is used, $\varepsilon_{g}$ decreases 
linearly as $\epstol$ is reduced, for both studied cases.
However, if different statistical seeds are used,
one can note a limit behavior where $\varepsilon_{g}$ does not decrease if 
$\epstol$ is reduced. This saturation in $\varepsilon_{g}$ value
($\sim 10^{-2}$) indicates that a decrease in $\epstol$ value 
below $10^{-3}$ is not effective in cases where the statistical
seeds are different.

\begin{figure*}[ht!]
    \centering
    \subfigure[Evolution of $\rerror{ \dless{\mean{T}} }$ for case~3.]
    {\includegraphics[scale=0.35]{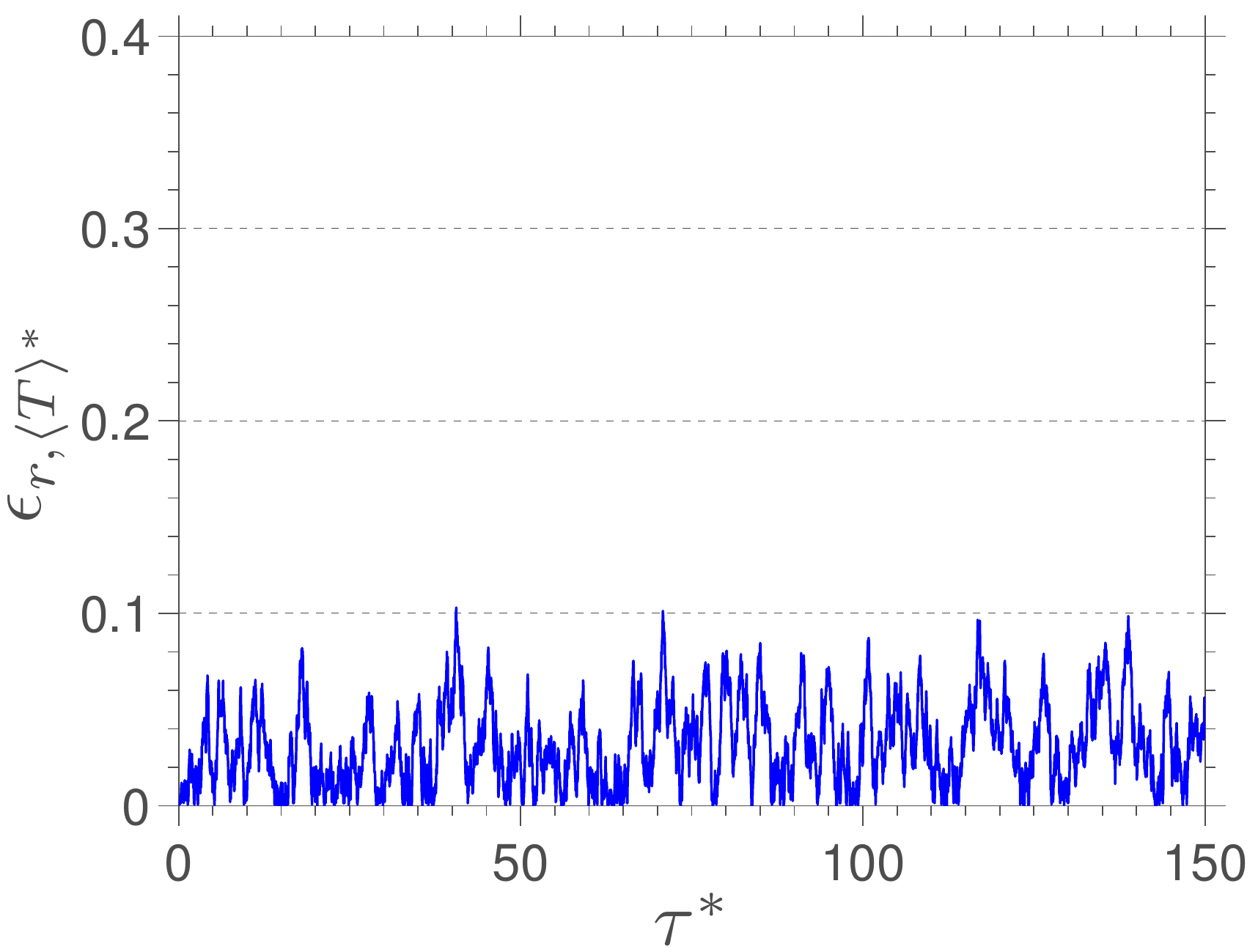}}
    \subfigure[Evolution of $\rerror{ \mean{Y_{OH}} }$ for case~3.]
    {\includegraphics[scale=0.35]{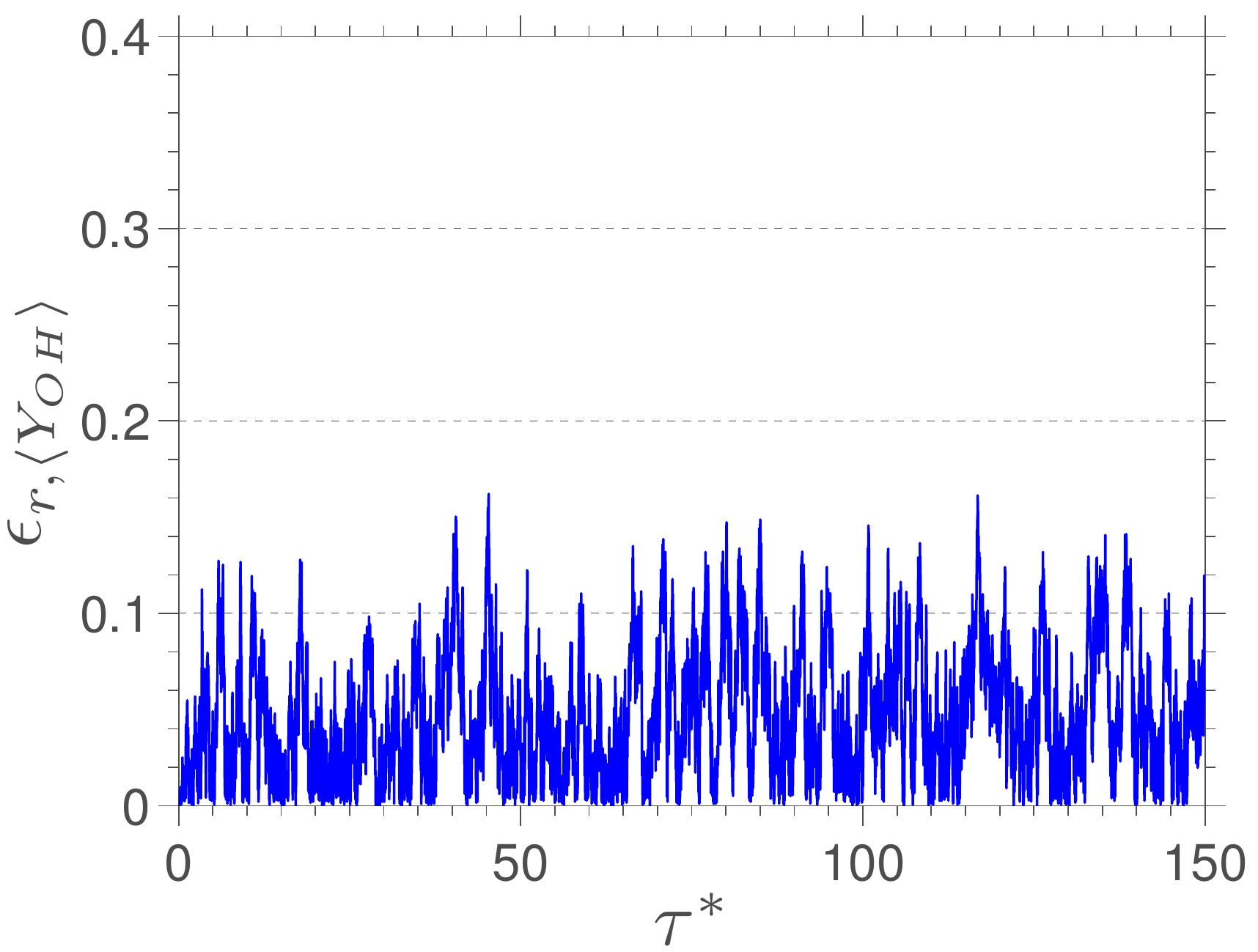}}\\
    \subfigure[Evolution of $\rerror{ \dless{\var{T}} }$ for case~3.]
    {\includegraphics[scale=0.35]{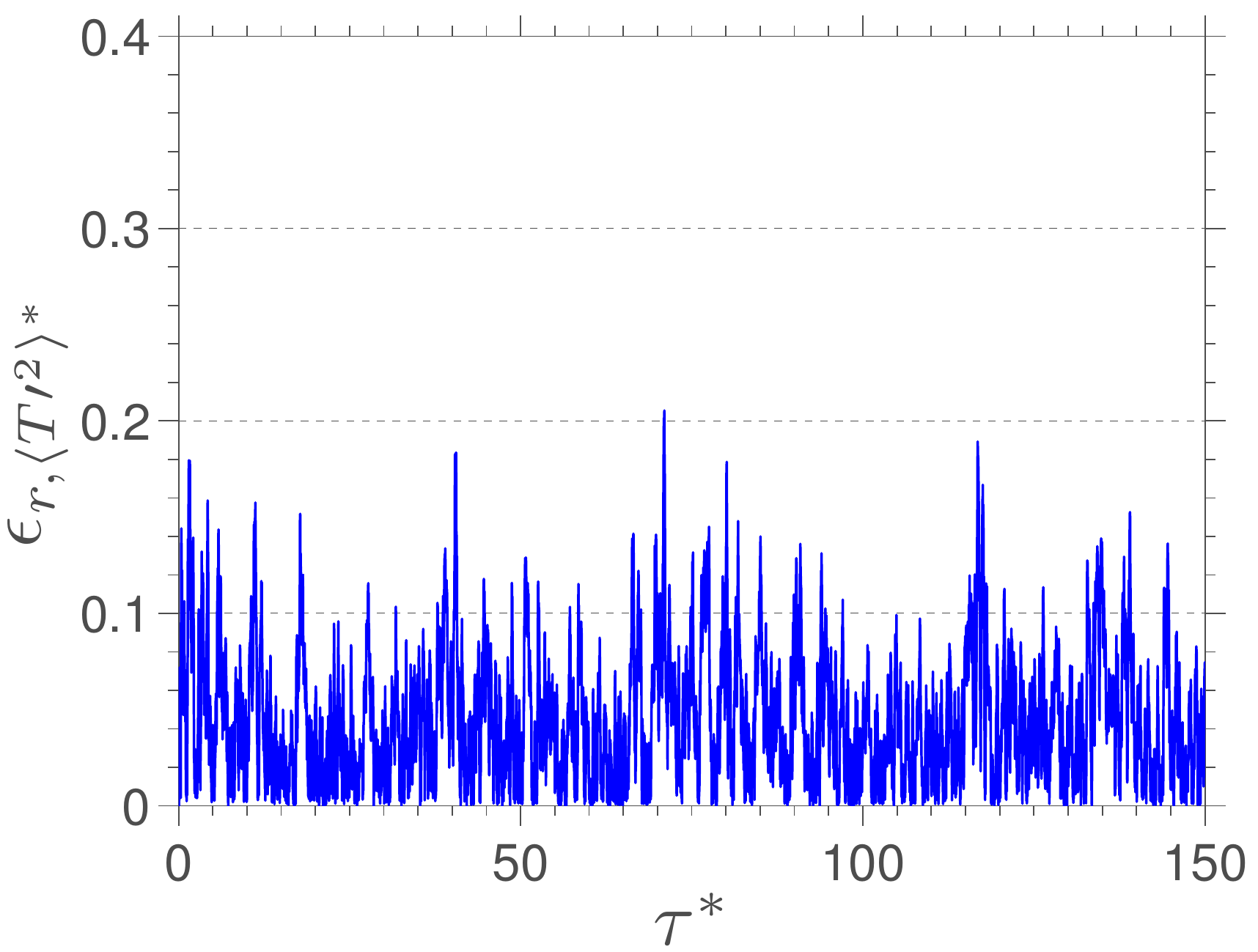}}
    \subfigure[Evolution of $\rerror{ \var{Y_{OH}} }$ for case~3.]
    {\includegraphics[scale=0.35]{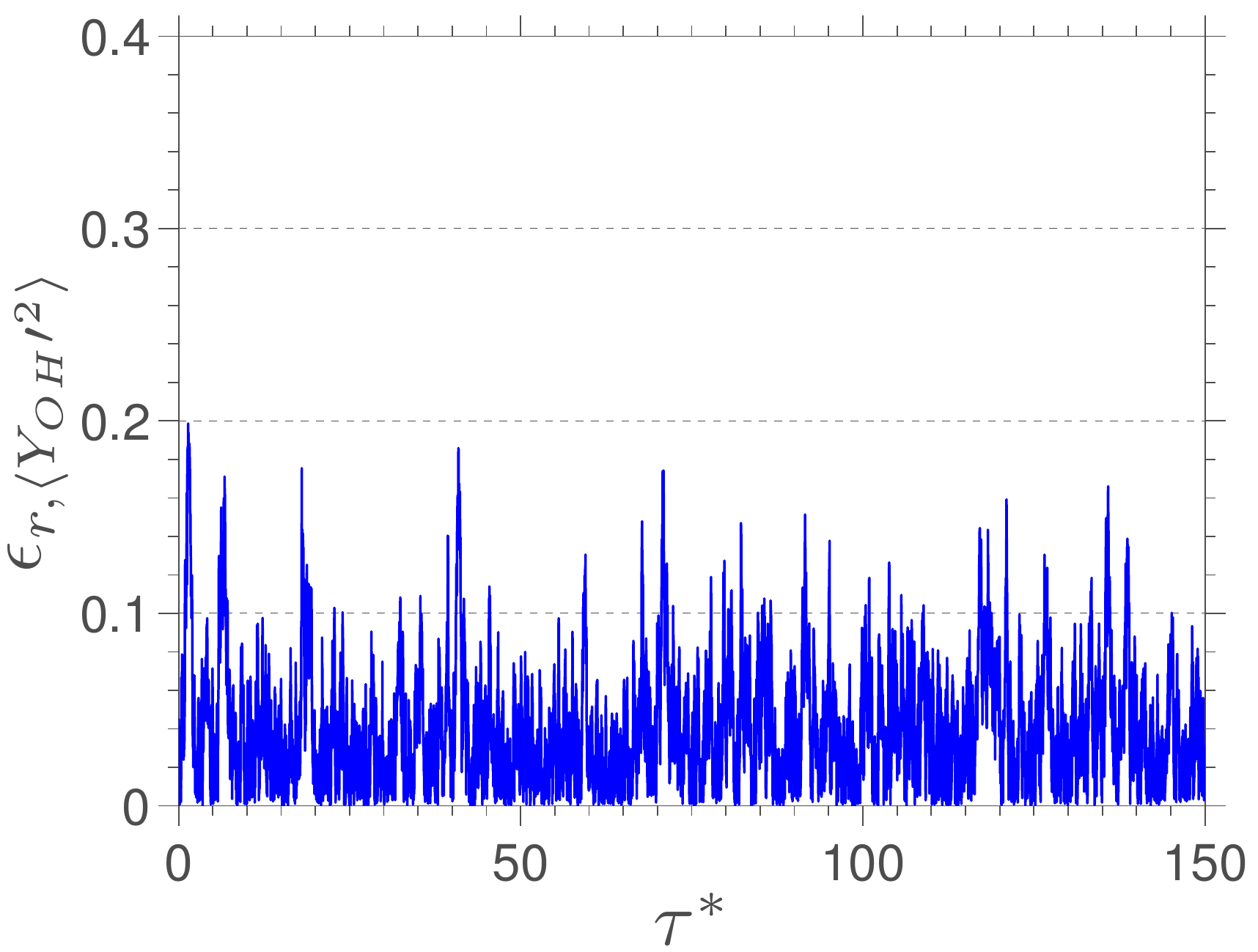}}\\
    \caption{Evolution of the relative local error of the ensemble average
    mean and the ensemble average variance of the reduced temperature and $OH$ mass fraction.}
    \label{case3_error_T_OH_mean_var_vs_t}
\end{figure*}

The evolution of the relative local error of the statistical moments 
of $\dless{T}$ and $Y_{OH}$, for case~3, are presented in
Figure~\ref{case3_error_T_OH_mean_var_vs_t}.
Compared with the results of case~1, the errors of case~3 
present more oscillations.  Furthermore, for this case,
$\varepsilon_{g} = 2.4 \times 10^{-3}$ is an order of 
magnitude larger than the values obtained
in cases~1~and~2, also using $\epstol=10^{-3}$.
Since different seeds are used for DI and ISAT calculations
in case~3, large errors in the results are to be expected.

\subsection{Analysis of ISAT performance}

\begin{figure*}[ht!]
    \centering
    \subfigure[Case~1.]
    {\includegraphics[scale=0.35]{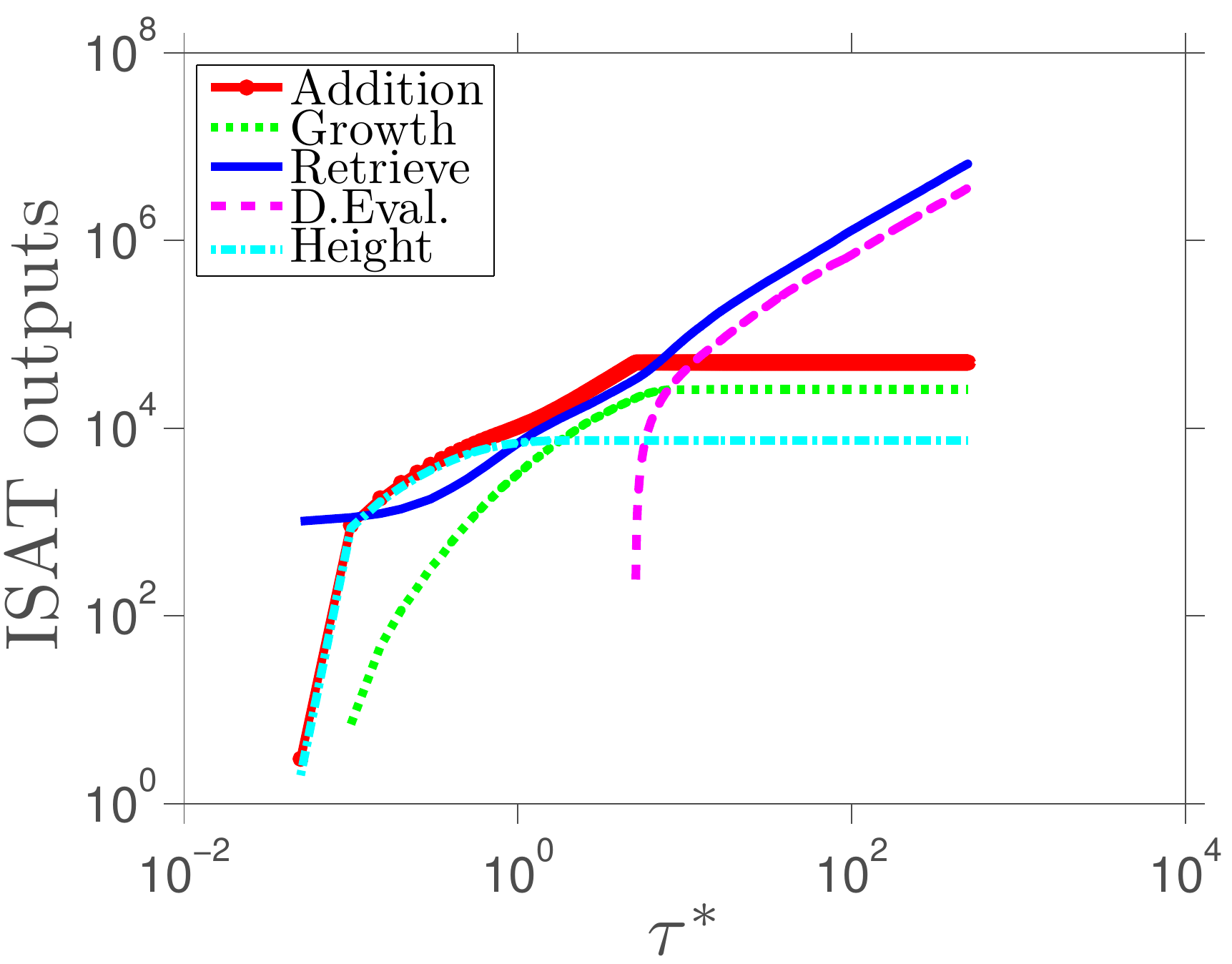}}
    \subfigure[Case~2.]
    {\includegraphics[scale=0.35]{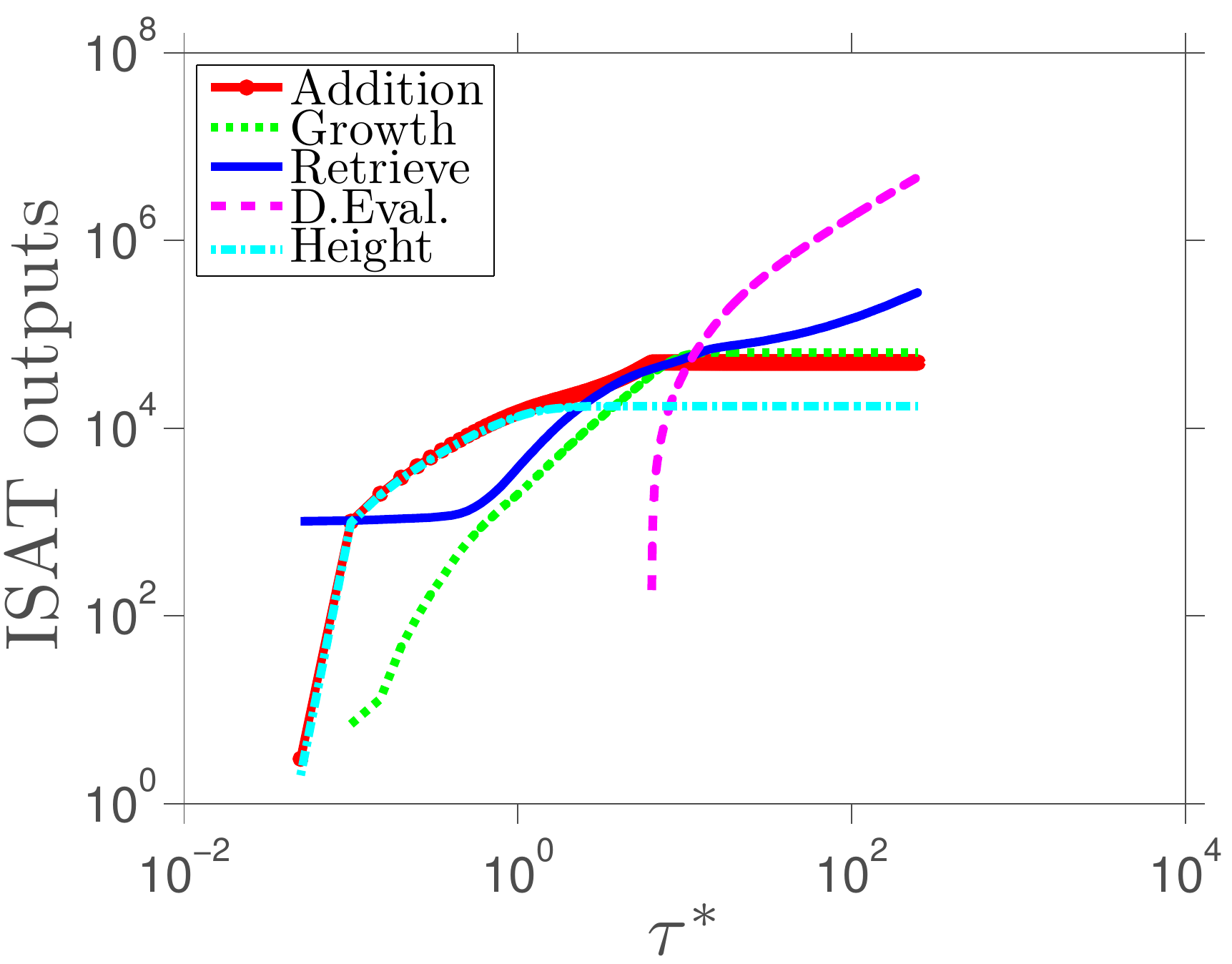}}
    \caption{Evolution of ISAT algorithm outputs and of the height of ISAT
    binary search tree.}
    \label{pmsr_CO_O2_isat_out_vs_t}
\end{figure*}

\begin{figure*}[ht!]
    \centering
    \subfigure[Case~1.]
    {\includegraphics[scale=0.35]{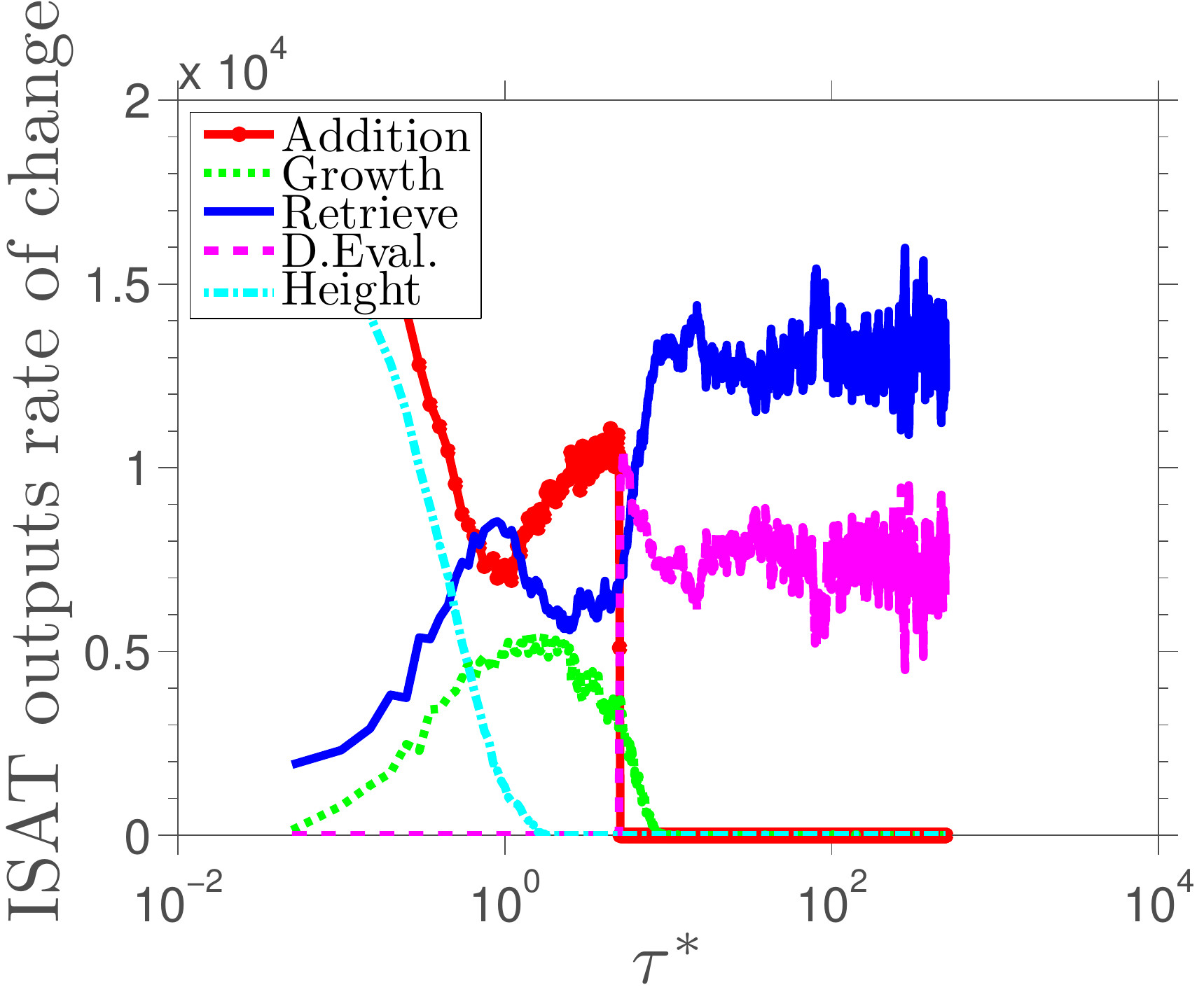}}
    \subfigure[Case~2.]
    {\includegraphics[scale=0.35]{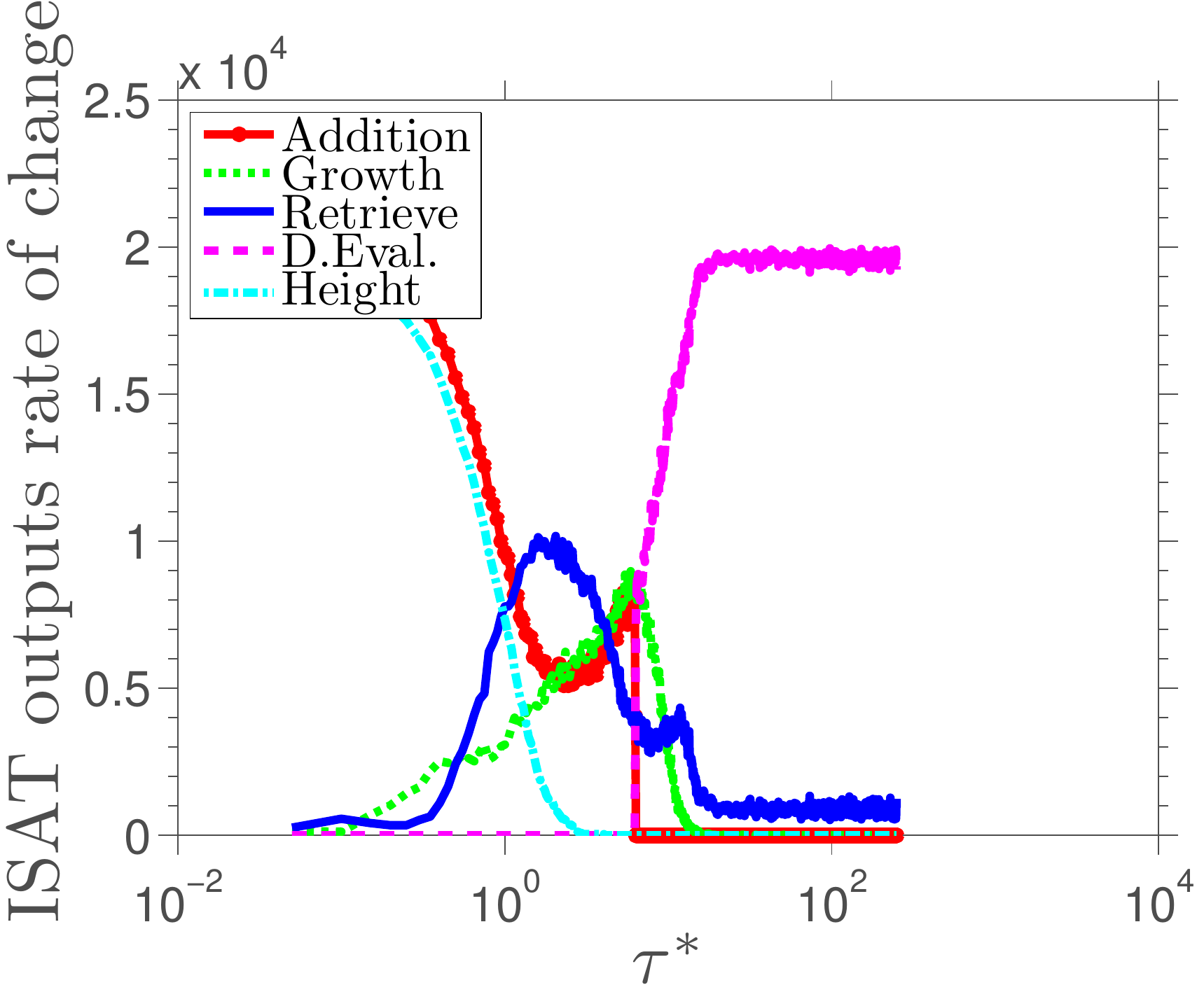}}
    \caption{Evolution of the rates of change of each ISAT algorithm outputs
    and of the height of ISAT binary search tree.}
    \label{pmsr_CO_O2_isat_out_dot_vs_t}
\end{figure*}

The comparison of time evolution of ISAT outputs
(number of additions, growths, retrieves, and direct evaluations)
and of  binary search tree height, as well as the corresponding 
rates of change, for cases~1~and~2, are presented in
Figures~\ref{pmsr_CO_O2_isat_out_vs_t}~and~\ref{pmsr_CO_O2_isat_out_dot_vs_t}.
A first important observation is that the number of additions in both cases
reaches the maximum prescribed value in the binary search tree of $50\unit{k}$.
As a consequence of binary search tree saturation, the additions curve 
reaches a steady state after $5.4$ and $1.2$ residence times ($\tau_r$), 
in the first and second cases, respectively. This indicates that ISAT table is 
saturated earlier when mixing is faster ($\tau_m / \tau_r$ small).

Figures~\ref{pmsr_CO_O2_isat_out_vs_t}~and~\ref{pmsr_CO_O2_isat_out_dot_vs_t}
also show the evolution of binary tree height, which reaches steady state 
$1.8~\tau_r$ in the first case and $0.4~\tau_r$ in the second case.
It is also noteworthy that, in both cases, the tree height is an order of
magnitude smaller than the total number of entries in the tree 
($\sim17\unit{k}$ in case~1 and $\sim7\unit{k}$ in case~2). 
This difference between height and total entries in data table ensures 
the efficiency of search process for a new query, which could be performed 
up to three and seven times faster in cases~1~and~2, respectively, 
than a vector search.

In the first case, 
Figures~\ref{pmsr_CO_O2_isat_out_vs_t} and \ref{pmsr_CO_O2_isat_out_dot_vs_t}
show that the number of growths presents a sharp rate of change
around $\tau_r$, whereas, in the second case, this occurs around $5~\tau_r$. 
In both cases, growth steady state occurs after $10~\tau_r$.
During both simulations the number of growths is always smaller than 
additions. This indicates that the desirable massive increase of the 
ellipsoids of accuracy, to improve the estimate of the region of accuracy, 
is not observed. This behavior might circumstantial to the 
reaction system of the carbon monoxide considered, since, due to its simplicity 
(only three reactions), a small part of the realizable region should be
assessed during the course of the calculation.

The Figure~\ref{pmsr_CO_O2_isat_out_vs_t} shows that, 
after tree saturation occurs, the number of retrieves, and direct evaluations 
exceed the number of additions in both cases. In case 1 there is a 
higher occurrence of retrieve events, whereas in case 2 
direct evaluations prevail. The number of retrieves exhibits a linear 
limit behavior in both cases. The ISAT behavior for the second case 
reflects the fact that  the binary tree of this case is poor, i.e., 
contains too few compositions in the region accessed by the calculation.
As a consequence, the number of direct evaluations vastly outnumbers
ISAT operations. This explains the error behavior observed in 
section~\ref{analysis_pmsr_accur}.

When ISAT technique is more efficient than DI procedure, 
the computational time spent by ISAT is smaller than 
computational time spent by DI. The computational time 
spent by ISAT is the sum of the computational time spent 
at each of its possible outputs. Therefore, the efficiency
condition can be stated as

\begin{equation}
    n_{A} \tau_{A} + n_{G} \tau_{G} + n_{R} \tau_{R} + n_{DE} \tau_{DE} 
    < n_{DI} \tau_{DI},
    \label{effic_cond1}
\end{equation}

\noindent
where 
$n_{A}$, $n_{G}$, $n_{R}$, $n_{DE}$, $n_{DI}$
are the number of additions, growths,
retrieves, direct evaluations, and direct integrations,
respectively, whereas
$\tau_{A}$, $\tau_{G}$, $\tau_{R}$, $\tau_{DE}$, and $\tau_{DI}$
are the corresponding average duration at each operation.

Assuming that

\begin{equation}
    \frac{\tau_{G}}{\tau_{DI}} \approx 1,
    \qquad
    \frac{\tau_{DE}}{\tau_{DI}} \approx 1,
    \qquad \mbox{and} \qquad
    \frac{\tau_{R}}{\tau_{DI}} \ll 1.
\end{equation}

\noindent
it is possible to show \cite{cunha2010} that a necessary (but not sufficient)
condition for a calculation using ISAT to be faster than the same calculation using DI is

\begin{equation}
    \frac{n_{R}}{n_{A}} >
    \frac{\tau_{A}}{\tau_{DI}} - 1.
    \label{add_ret_relation}
\end{equation}

For $CO/O_{2}$ mixtures estimates lead to
$\tau_{A} / \tau_{DI} \approx 10$, \cite{cunha2010}. 
It is possible to see in
Figure~\ref{pmsr_CO_O2_isat_out_vs_t} that, for case~1, 
$n_{R}$ approximately exceeds $n_{A}$ by a factor of $130$,
whereas in case~2 this factor is only $6$. 
Therefore, as $\tau_{A} / \tau_{DI} < 10$ in case~2, 
ISAT calculations are not expected to be faster than DI procedure.

As can be seen in Table~\ref{pmsr_CO_O2_cpu_time}, 
where a comparison of computational time is shown, 
cases~1~and~2, for $\epstol = 10^{-3}$ are computed 
using DI in $4.0~\unit{ks}$ and $2.0~\unit{ks}$
respectively, whereas with the use of ISAT, the same cases spent
$2.3~\unit{ks}$ and $2.1~\unit{ks}$, respectively.
Speed-up factors of $1.7$ (case~1) and $1.0$ (case~2) 
are obtained, where the \emph{speed-up factor} is defined as the 
ratio between the computational time spent by DI and 
the computational time spent by ISAT. 
This table also allows to compare the computational time spent
by DI and ISAT for different values of error tolerance.
An increase in processing time is obtained as ISAT error tolerance is reduced, 
which is to be expected, given the fact that lower values of $\epstol$
correspond to a smaller region of accuracy. Indeed, as $\epstol$ is
decreased, it is less likely that ISAT returns a retrieve, which is ISAT output 
with lower computational cost. In case~1, for all values of $\epstol$, 
ISAT technique offers an advantage, in terms of computational time,
when compared to the process of direct integration, reducing on average the 
computational time in 42\%. On the other hand, in case~2, no reduction in 
computational time is seen. As discussed above, this behavior is natural, 
once the necessary condition for efficiency of the algorithm is not reached.

\begin{table*}[ht!]
	\centering
	\caption{Comparison between the computational time
	spent by DI and ISAT and the corresponding 
	speed-up factors.}
	\input{tabs/pmsr_CO_O2_cpu_time_50k.tab}
	\label{pmsr_CO_O2_cpu_time}
\end{table*}

Aiming to analyze the asymptotic behavior of ISAT speed-up,
case~1 has also been simulated for 50,000 time steps. This computation
with ISAT spends $204.7~\unit{ks}$, whereas if DI were to be used the same
calculation would spend $\sim1024.0~\unit{ks}$ (speed-up factor of 5.0).
Therefore, in this asymptotic case, ISAT spends approximately 
80\% less time than DI.
The pioneer work of Pope (1997) \cite{pope1997p41} reports an 
asymptotic speed-up factor of 1000, but this impressive factor
has not been observed in the study developed here.

\begin{figure*}[ht!]
    \centering
    \subfigure[Outputs for case~3.]
    {\includegraphics[scale=0.35]{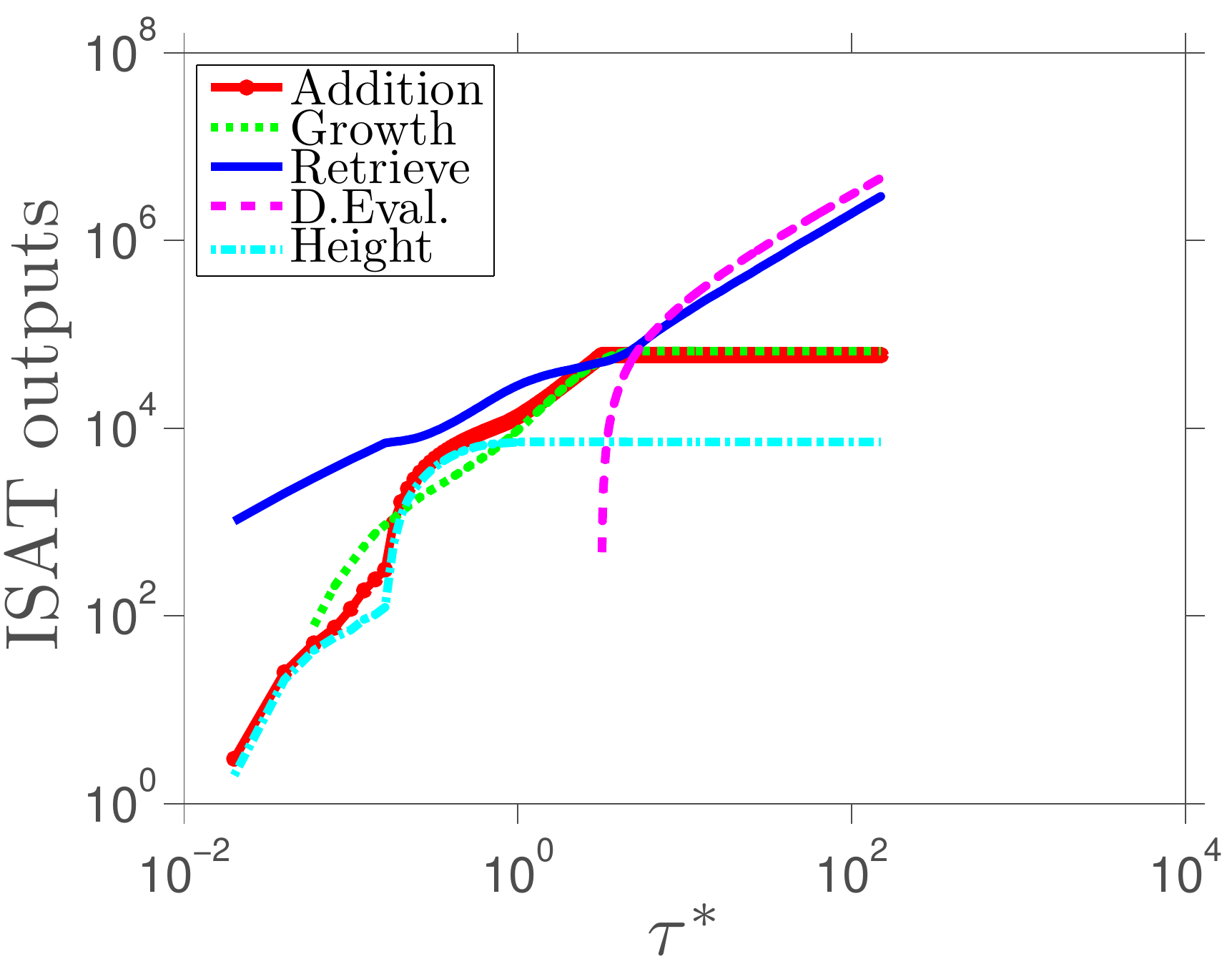}}
    \subfigure[Outputs rates of change for case~3.]
    {\includegraphics[scale=0.35]{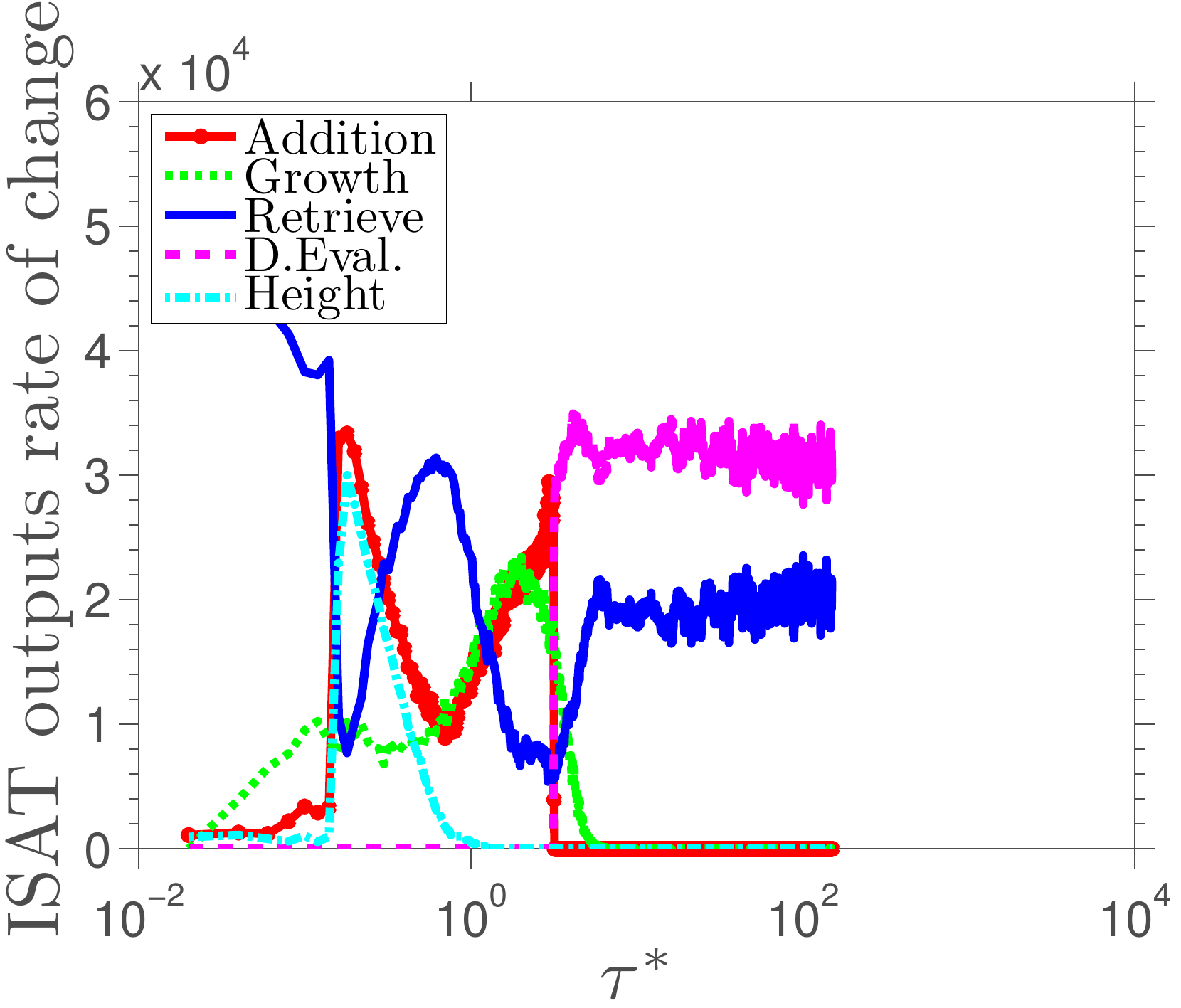}}
    \caption{Evolution of ISAT algorithm outputs, height of ISAT
    binary search tree, and the corresponding rates of change.}
    \label{case3_isat_out_vs_t}
\end{figure*}

The evolution of ISAT algorithm outputs, 
ISAT binary search tree height, and the corresponding 
rates of change, for case~3, are presented in
Figure~\ref{case3_isat_out_vs_t}.
As in cases~1~and~2 the number of additions reaches the maximum 
allowed value in the binary search tree, $60\unit{k}$ for this case.
The additions curve reaches a steady state after 
$\dless{\tau} = 3.2$. The binary search tree maximum height 
for case~3 is $\sim7\unit{k}$ as in case~2. Here the steady state of 
the tree height is observed at $\dless{\tau} = 1.28$, as can be seen in 
Figure~\ref{case3_isat_out_vs_t}.
The behavior of the number of growths is quite similar to
that of case~1, where the greater rate of change occurs near
$\dless{\tau} = 2$ but, now, the steady state is reached before 
$\dless{\tau} = 10$.

For case~3, nevertheless, the number of growths is not always smaller 
than the number of additions. Initially the growths exceed additions 
by nearly an order of magnitude. After $\dless{\tau} = 0.2$
the additions exceeded the number of growths, only to be overcome 
again once steady state is reached. This behavior indicates that 
the ellipsoids of accuracy show a considerable period of adaptation,
which allows to access, with precision, a larger portion of the 
realizable region.

As in the previous cases, after tree saturation, the number of 
retrieves and direct evaluations exceed the number of additions.
The number of direct evaluations overcomes total number of retrieves
in approximately 60\%, which is not negligible, but less than
that observed in case~2. This indicates
that the binary search tree covers a significant portion of the 
realizable region in the composition space.
From Figure~\ref{case3_isat_out_vs_t} one can estimate 
$\tau_{A} / \tau_{DI}  \approx 50$ in order that ISAT 
technique be more efficient than DI.

For case~3, the computational time spent by DI and ISAT 
are $689~\unit{ks}$ and $455~\unit{ks}$,
respectively. Therefore, a speed-up of 1.5 is observed. 
In this case, the effect of other values of ISAT error 
tolerance has not been investigated. 
This PMSR has 1024 particles and its evolution is 
computed during $\dless{\tau} = 150$.
Thus, the system of governing equations defined by 
Eq.(\ref{eq_reaction}) is solved $7,680,000$ times. 
In this case ISAT technique allows to save 34\%, in terms 
of computational time. For problems that require solving 
Eq.(\ref{eq_reaction}) several times, one could speculate 
that ISAT technique would provide an even better 
performance improvement, since more retrieves are expected 
when compared with the more costly direct integration.

\subsection{Analysis of ISAT memory usage}

The studied cases~1~and~2  are both modeled
by a reaction mechanism with 4~species and use a binary
search tree with 50,000 entries for ISAT simulations.
These parameters lead to a memory consumption by
ISAT algorithm of approximately $40~\unit{Mbytes}$,
which is very small when compared with the available memory
in the workstation used.

On the other hand, case~3 is modeled by a reaction 
mechanism with 53~species and use a binary
search tree with 60,000 entries. This case uses approximately 
$3.3~\unit{Gbytes}$ of memory. It is noteworthy that this value
of memory usage is not negligible, when compared to the total 
memory available on the workstation. Thus, for practical purposes, 
this tree is the largest one that may be used to simulate a PMSR 
with this methane reaction mechanism and 1024 particles.
This underscores what is perhaps the greatest shortcoming of ISAT 
technique, its huge expense of memory.

\section{Concluding remarks}

This work assessed characteristics of a PMSR reactor through
the numerical simulation of reactor configurations with simple 
($CO/O_{2}$) and complex ($CH_4$/air) reaction mechanisms
and also investigated ISAT technique as an option to evaluate a 
computational model with detailed thermochemistry.
For this purpose, the accuracy, performance, and 
memory usage of the corresponding technique were analyzed through the 
comparison of ISAT results with a reference result, obtained from 
direct integration of PMSR model equation. The studied cases analysis 
showed that ISAT technique has a good accuracy from a global 
point of view. Also, it was possible to identify the statistical seed 
effect on the control of absolute global error, which decreases 
monotonically as ISAT error tolerance was reduced, when the same 
seed is used for ISAT and DI calculations. On the other hand, 
when different seeds were used, a limit value for ISAT error tolerance 
was observed. In terms of performance, ISAT technique allows to reduce 
the computational time of the simulations in all cases studied. 
For $CO/O_{2}$ cases, a speed-up of up to 5.0 was achieved, 
whereas for $CH_4$/air, the algorithm allowed to save 34\% in terms 
of computational time. Moreover, ISAT technique presented the 
desired feature of speed-up factor increase with the complexity of analyzed 
system. Regarding the memory usage, ISAT technique showed to be very 
demanding. This work will be extended by coupling ISAT technique to the 
hybrid LES/PDF model by 
\cite{andrade2011,vedovoto2011p85,andrade2009,vedovoto2011}
for description of turbulent combustion, when a detailed reaction 
mechanism would allow a better description of combustion.
In this context, ISAT could be a viable option that may be able to decrease,
to an acceptable level, the simulation time.

\begin{acknowledgements}
The authors acknowledge the support awarded to this research from
CNPq, FAPERJ, ANP and Brazilian Combustion Network.
The authors are indebted to Professor Guenther Carlos Krieger Filho from 
Universidade de S\~{a}o Paulo (USP), for providing the code that served 
as example to the code developed in this work, and for his hospitality 
during the authors' visit to USP. This work was performed while the second
author was on leave from Institut Pprime, CNRS, France.
\end{acknowledgements}

\bibliographystyle{spbasic}      
\bibliography{references}

\end{document}

%% file: macros.tex
\newcommand{\unit}[1]{\ensuremath{\mathrm{#1}}}

\newcommand{\define}[1]{\emph{#1}\index{#1}}

\newcommand{\epstol}{\ensuremath{\varepsilon_{tol}}}

\newcommand{\transp}[1]{\ensuremath{#1^{\mathtt{T}}}}

\newcommand{\dless}[1]{\ensuremath{#1^{\ast}}}

\newcommand{\order}[1]{\ensuremath{\mathcal{O}( #1 )}}

\newcommand{\rerror}[1]{\ensuremath{\varepsilon_{_{r}, #1 } }}

\newcommand{\funcM}[1]{\ensuremath{\mean{ #1 }_{M}}}

\newcommand{\mean}[1]{\ensuremath{\left \langle #1 \right \rangle}}

\newcommand{\var}[1]{\ensuremath{\left \langle #1'^{2} \right \rangle}}

\newcommand{\ceil}[1]{\ensuremath{\textsl{ceil} \left( #1 \right)}}



%% file: tabs/pmsr_errors.tab
%
%
%

\begin{tabular}{lrrrr}
\toprule
			   & \multicolumn{2}{c}{Equal Seeds} & \multicolumn{2}{c}{Different Seeds} \\
\cmidrule(r){2-3} \cmidrule(r){4-5}
	$\psi$		         & Case 1 & Case 2 & Case 1 & Case 2 \\
\cmidrule(r){2-2} \cmidrule(r){3-3} \cmidrule(r){4-4} \cmidrule(r){5-5}
	$\dless{\mean{T}}$       & $~1.0~\%$ & $~0.0~\%$ &~~8.1~\% &~~5.3~\% \\
	$\mean{Y_{O}}$           & $~1.2~\%$ & $~0.1~\%$ &~11.7~\% &~~1.3~\% \\
	$\dless{\var{T}}$        & $~1.0~\%$ & $~0.5~\%$ &~~6.6~\% &~10.1~\% \\
	$\var{Y_{O}}$            & $~0.9~\%$ & $~0.5~\%$ &~~7.0~\% &~~6.1~\% \\
\bottomrule
\end{tabular}

%% file: tabs/pmsr_CO_O2_cpu_time_50k.tab
%
%

\begin{tabular}{lcccc}
\toprule
		  & \multicolumn{2}{c}{Case 1} & \multicolumn{2}{c}{Case 2} \\
\cmidrule(r){2-3} \cmidrule(r){4-5}
	$\epstol$ & time spent ($\unit{ks}$) & speed-up    & time spent ($\unit{ks}$) & speed-up \\
\midrule
	DI        & $\sim 4.0$               &             & $\sim 2.0$               & \\
	$10^{-2}$ & $\sim 1.9$               & $\sim 2.1$  & $\sim 2.1$               & $\sim 1.0$\\
	$10^{-3}$ & $\sim 2.3$               & $\sim 1.7$  & $\sim 2.1$               & $\sim 1.0$\\
	$10^{-4}$ & $\sim 2.4$               & $\sim 1.7$  & $\sim 2.2$               & $\sim 0.9$\\
	$10^{-5}$ & $\sim 2.5$               & $\sim 1.6$  & $\sim 2.2$               & $\sim 0.9$\\
	$10^{-6}$ & $\sim 2.6$               & $\sim 1.5$  & $\sim 2.2$               & $\sim 0.9$\\
\bottomrule
\end{tabular}